\documentclass[journal, onecolumn]{IEEEtran}
\renewcommand{\baselinestretch}{2.0}

\ifCLASSINFOpdf
\else
\fi

\include{epsf}
\include{epsfrot}
\usepackage[dvips]{graphicx}
\usepackage{subcaption}
\usepackage[skip=2pt,font=small]{caption}
\usepackage{epsfig}
\usepackage{afterpage}
\usepackage{graphicx}
\usepackage[centertags]{amsmath}
\usepackage{amsfonts}
\usepackage{amssymb}
\usepackage{amsthm}
\usepackage{nccmath}
\usepackage{color}
\usepackage{algorithm}
\usepackage{algpseudocode}
\usepackage{setspace}
\let\Algorithm\algorithm
\renewcommand\algorithm[1][]{\Algorithm[#1]\setstretch{1.2}}

\def \be {\begin{eqnarray}}
\def \ee {\end{eqnarray}}
\def \bc {\begin{center}}
\def \ec {\end{center}}
\def\bfig { \begin{figure} }
\def\efig { \end{figure} }
\def\bit { \begin{itemize} }
\def\eit { \end{itemize} }
\def\benum { \begin{enumerate} }
\def\eenum { \end{enumerate} }

\begin{document}
\title{Multivariate Fronthaul Quantization for \\ Downlink C-RAN}

\author{Wonju~Lee,~\IEEEmembership{Student~Member,~IEEE,}
        Osvaldo~Simeone,~\IEEEmembership{Fellow,~IEEE,}
        Joonhyuk~Kang,~\IEEEmembership{Member,~IEEE,}
        and~Shlomo~Shamai~(Shitz),~\IEEEmembership{Fellow,~IEEE}% <-this % stops a space
\thanks{W. Lee and J. Kang are with the Department of Electrical Engineering, Korea Advanced Institute of Science and Technology (KAIST),
Daejeon, 305-701, South Korea (e-mail: wonjulee@kaist.ac.kr, jhkang@ee.kaist.ac.kr).}% <-this % stops a space
\thanks{O. Simeone is with the Center for Wireless information Processing (CWiP),
Department of Electrical and Computer Engineering, New Jersey Institute of Technology (NJIT),
Newark, NJ 07102, USA (e-mail: osvaldo.simeone@njit.edu).
The work of O. Simeone was partially supported by U.S. NSF under grant CCF-1525629.}
\thanks{S. Shamai (Shitz) is with the Department of Electrical Engineering,
Technion, Haifa, 32000, Israel (e-mail:  sshlomo@ee.technion.ac.il).
The work of S. Shamai was supported by the Israel Science Foundation and the S. AND N. grand research fund.}}% <-this % stops a space

\markboth{Journal of \LaTeX\ Class Files,~Vol.~X, No.~XX, July~2016}%
{Shell \MakeLowercase{\textit{et al.}}: Bare Demo of IEEEtran.cls for Journals}

\maketitle

\begin{abstract}
The Cloud-Radio Access Network (C-RAN) cellular architecture relies on the transfer of complex baseband signals to and from a central unit (CU) over digital fronthaul links to enable the virtualization of the baseband processing functionalities of distributed radio units (RUs).
The standard design of digital fronthauling is based on either scalar quantization or on more sophisticated point-to-point compression techniques operating on baseband signals.
Motivated by network-information theoretic results, techniques for fronthaul quantization and compression that improve over point-to-point solutions by allowing for \emph{joint} processing across multiple fronthaul links at the CU have been recently proposed for both the uplink and the downlink.
For the downlink, a form of joint compression, known in network information theory as \emph{multivariate compression}, was shown to be advantageous under a non-constructive asymptotic information-theoretic framework.
In this paper, instead, the design of a practical symbol-by-symbol fronthaul quantization algorithm that implements the idea of multivariate compression is investigated for the C-RAN downlink.
As compared to current standards, the proposed multivariate quantization (MQ) only requires changes in the CU processing while no modification is needed at the RUs.
The algorithm is extended to enable the joint optimization of downlink precoding and quantization, reduced-complexity MQ via successive block quantization, and variable-length compression.
Numerical results, which include performance evaluations over standard cellular models, demonstrate the advantages of MQ and the merits of a joint optimization with precoding.
\end{abstract}

\IEEEpeerreviewmaketitle

\section{Introduction}
Cloud Radio Access Network (C-RAN) refers to the centralization of base station functionalities by means of cloud computing. 
It is universally accepted as one of the key technologies for $5$G systems due to its significant advantages in terms of lower expenses, flexibility and enhanced spectral efficiency.
In particular, C-RAN enables the implementation of coordinated multipoint (CoMP) schemes across all the ratio units (RUs) connected to the central processing unit (CU) in the ``cloud'' \cite{China}.
The main obstacle to the realization of the promises of C-RAN resides in the restrictions on the capacity and latency of the so-called fronthaul links that provide connectivity between RUs and the CU (see \cite{Checko} for a review and Fig. \ref{fig:SM} for an illustration).
The standard design of digital fronthauling, that is, of the transmission of digitized baseband complex samples on the fronthaul links, is based on either scalar quantization, as in the Common Public Radio Interface (CPRI) standard,
or on more sophisticated \emph{point-to-point} compression techniques operating on baseband signals \cite{Samardzija}-\cite{Si}.
This means that a \emph{separate} quantizer, and possibly compressor, is implemented for each fronthaul link and hence for each RU connected to the CU \cite{Checko}.

Motivated by network-information theoretic results, techniques for fronthaul quantization/compression that improve over point-to-point solutions
by allowing for joint processing across multiple fronthaul links at the CU have been studied for both the uplink and the downlink (see \cite{Park:SPM} and references therein).
In particular, for the uplink, distributed source coding, which can be implemented by means of Wyner-Ziv coding, was demonstrated to yield significant performance gains by leveraging \emph{joint decompression} at the CU \cite{Park:SPM}\cite{Zhou}.
In a dual fashion, for the downlink, a form of \emph{joint compression}, known in network information theory as \emph{multivariate compression}, was shown to be advantageous in \cite{Park:TSP}.
In both cases, the design and performance gains of joint decompression or joint compression depend on the density of the deployment of RUs in the network and rely on some degree of channel state information (CSI) regarding the uplink or downlink channels at the CU.

The results reviewed above concerning the advantages of joint fronthaul processing at the CU were derived under a non-constructive asymptotic information-theoretic framework that assumes long coding blocks and only asserts the existence of coding schemes that achieve given performance bounds.
In this paper, instead, we consider the downlink of a C-RAN and investigate the design of a quantization algorithm that implements the idea of multivariate compression put forth in \cite{Park:SPM} and is hence referred to as Multivariate Quantization (MQ).
We specifically focus on a practical implementation that operates on a symbol-by-symbol basis.
Hence, as compared to conventional CPRI, MQ only requires changes in the CU processing, whereby quantization is performed jointly, rather than separately, on the baseband signals intended for multiple RUs, while no modification is needed at the RUs, which still perform scalar decompression.
Furthermore, following the theoretical insights of \cite{Park:TSP} on the benefits of joint optimization of fronthaul compression and downlink precoding, we propose an algorithm for the joint design of MQ and beamforming.
Next, to tackle the computational complexity issue of MQ, we propose a reduced-complexity MQ by means of successive block quantization.
Finally, we investigate the potential advantages of complementing scalar quantization with variable-length compression by designing an entropy-constrained version of MQ.
The proposed algorithms are based on iterative optimization approaches similar to the classical Lloyd-Max and Linde-Buzo-Gray algorithms \cite{Gray} along with successive convex approximation techniques \cite{Tao}.

The rest of the paper is organized as follows.
Sec. \ref{ch:SM} presents the system model and Sec. \ref{ch:MQ_principle} introduces with a simple example the key ideas behind MQ.
Note that the intuitive arguments in Sec. \ref{ch:MQ_principle} were not presented in \cite{Park:SPM} and provide fresh insights into the benefits and design of MQ.
Sec. \ref{ch:MQ_sep} describes the proposed algorithm design, and Sec. \ref{ch:MQ_JD} discusses the joint optimization of downlink precoding and MQ.
In turn, Sec. \ref{ch:sbMQ} handles the complexity occurred by MQ and proposes a reduced-complexity MQ via successive block quantization
and Sec. \ref{ch:MQ_VL} investigates the entropy-constrained MQ design.
Sec. \ref{ch:NR} shows numerical results and Sec. \ref{ch:MQ_LTE} demonstrate the system-level performance of MQ under standard cellular models.
Lastly, Sec. \ref{ch:Con} offers some final remarks.

\bfig[t]
\bc
\centering \epsfig{file=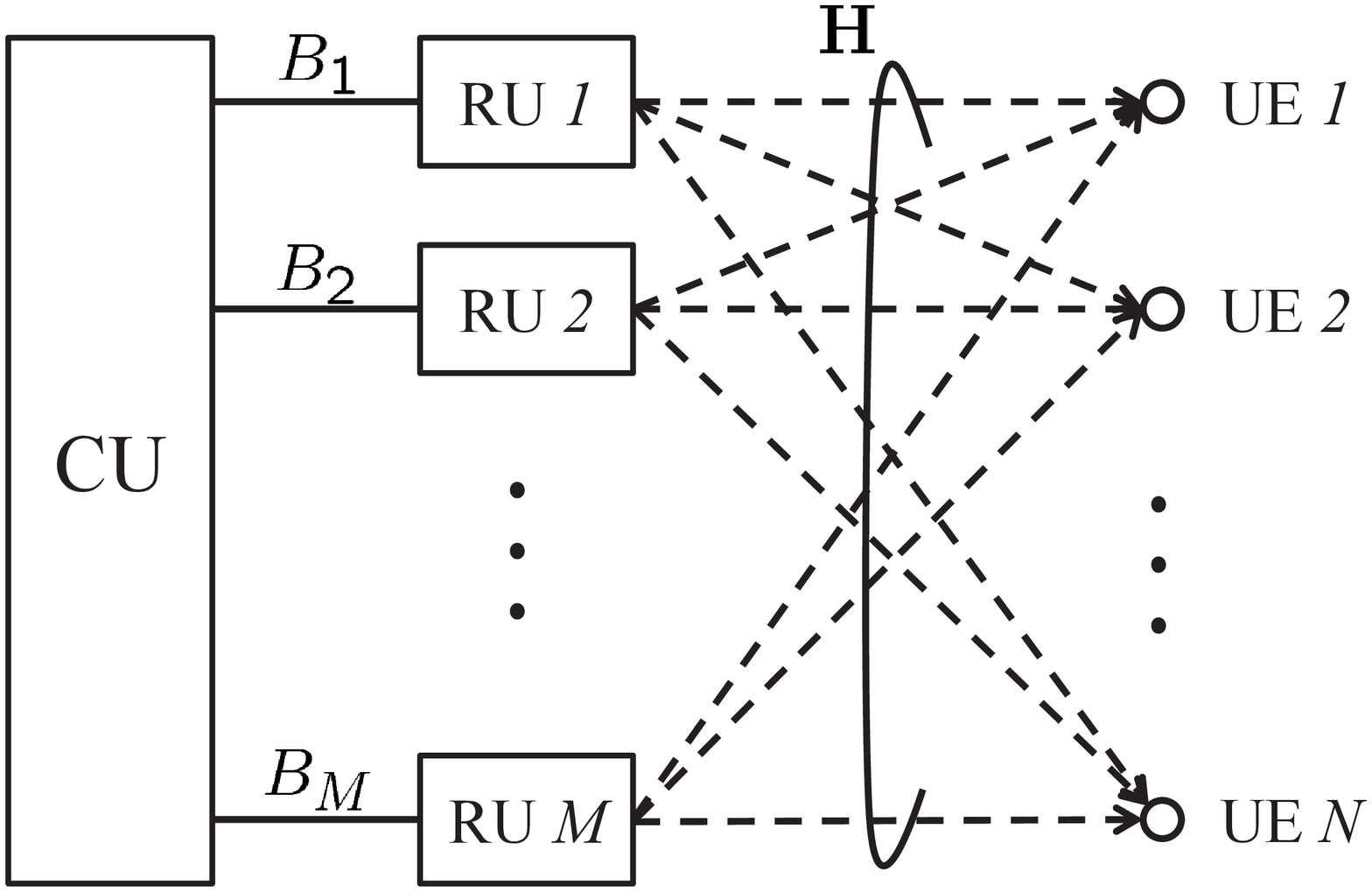, width=9cm, clip=}
\ec
\vspace*{-5mm}
\caption{Downlink C-RAN system with $M$ RUs and $N$ UEs. Each fronthaul link $i$ can carry $B_{i}$ bits per complex baseband sample.}
\label{fig:SM}
\efig

\section{System Model and Design Criterion} \label{ch:SM}
We consider the downlink of a C-RAN in which $M$ RUs cover an area with $N$ active user equipments (UEs), as illustrated in Fig \ref{fig:SM}.
Baseband processing for the $M$ RUs is carried out at a CU, hence enabling cooperative transmission, or CoMP in LTE jargon.
The CU transfers the baseband signals to each RU through fronthaul link with capacity $B_{i}$ as measured in bits per complex baseband sample, for $i=1,\dots,M$.

Let us define as $\mathbf{s}=[s_{1},\dots,s_{N}]^{T}$ the $N \times 1$ vector of complex information-bearing symbols at a given channel use, where $s_{k}$ is the symbol intended for UE $k$ that satisfies the normalization $E[|s_{k}|^{2}]=1$.
Each symbol $s_{k}$ may be taken from a finite constellation with uniform probability or it may be assumed to be distributed as a zero-mean complex Gaussian variable.
The latter case is often appropriate so as to obtain a modulation-independent solution or to account for OFDM transmission in the time domain by the law of large numbers as in e.g., \cite{Nieman}\cite{Grieger}.

In order to enable multi-user transmission, the information-bearing vector $\mathbf{s}$ is linearly precoded.
Assuming full CSI at the CU, this is done by means of an $M \times N$ precoding matrix $\mathbf{W}_\mathbf{H}=[\mathbf{w}_{\mathbf{H},1},\dots,\mathbf{w}_{\mathbf{H},N}]$,
where $\mathbf{w}_{\mathbf{H},k}$ is the beamforming, or precoding, vector for the signal $s_{k}$ intended for UE $k$.
The subscript $\mathbf{H}$ indicates the dependence of the precoding vectors on the channel matrix $\mathbf{H}$, to be introduced below.
Hence, the precoded signal $\mathbf{x}=[x_{1},\dots,x_{M}]^T$ is given as
\be \label{eq:precode_x}
\mathbf{x}=\sum_{k=1}^{N}\mathbf{w}_{\mathbf{H},k}s_{k}=\mathbf{W}_{\mathbf{H}}\mathbf{s}.
\ee

To satisfy the capacity limitations of the fronthaul links, the precoded signal $x_{i}$ for the RU $i$ is quantized to $B_{i}$ bits producing the signal $\hat{x}_{i}$.
The signal $\hat{x}_{i}$ is selected from a space $\hat{\mathcal{X}}_{i}$ cardinality $2^{B_{i}}$, which will be referred to as a codebook.
We define the quantized $M \times 1$ vector as $\hat{\mathbf{x}}=[\hat{x}_{1},\dots,\hat{x}_{M}]^T$ and assume the per-RU power constraint
\be \label{eq:per_power_cons}
E[\left|\hat{x}_{i}\right|^2]\leq 1
\ee
for all $i=1,\dots,M$. Each RU $i$ is assumed to be informed about the codebook $\hat{\mathcal{X}}_{i}$.
No additional information, such as CSI, is instead assumed at the RU.
Furthermore, codebooks are assumed to be updated only at the time scale of the variations of the long-term statistical properties of the channels $\mathbf{H}$.
As a result, RUs need to be informed about new codebooks only when the statistics of the channels, such as path-loss and shadowing, change significantly.

Since each RU $i$ transmits $\hat{x}_{i}$, the received signal at the UE $k$ can be written as
\be \label{eq:rec_Sig}
y_{k} &=& \sqrt{P}\mathbf{h}_{k}^{T}\hat{\mathbf{x}}+z_{k} \nonumber \\
&=& \sqrt{P}\mathbf{h}_{k}^{T}\mathbf{w}_{\mathbf{H},k}s_{k}+\sqrt{P}\mathbf{h}_{k}^{T}\left(\hat{\mathbf{x}}-\mathbf{w}_{\mathbf{H},k}s_{k}\right)+z_{k},
\ee
where $P$ is a dimensionless parameter that accounts for the transmitted power of the RUs;
$\mathbf{h}_{k}$ is the $M \times 1$ channel vector which is assumed to have a given distribution e.g., Rayleigh with possibly correlated entries;
and $z_{k}\sim\mathcal{CN}(0,1)$ is the additive Gaussian noise.
From (\ref{eq:rec_Sig}), the effective received signal to noise ratio (SNR) at the UE $k$ can be obtained as
\be \label{eq:effSNR}
\textrm{SNR}_{k}^{\textrm{eff}}=\frac{P E\left[\left|\mathbf{h}_{k}^{H}\mathbf{w}_{\mathbf{H},k}\right|^2\right]}{1+P E\left[\left|\mathbf{h}_{k}^{H}(\mathbf{w}_{\mathbf{H},k}s_{k}-\hat{\mathbf{x}})\right|^2\right]},
\ee
where the expectation is taken with respect to the distribution of the channel $\mathbf{H}=[\mathbf{h}_{1},\dots,\mathbf{h}_{N}]$ and of $\mathbf{s}$, and the second term in the denominator measures the power of the interference term in (\ref{eq:rec_Sig}) due to quantization.

We observe that quantization only affects the interference power in (\ref{eq:effSNR}) and hence optimal quantizers should minimize the powers $E[|\mathbf{h}_{k}^{H}(\mathbf{w}_{\mathbf{H},k}s_{k}-\hat{\mathbf{x}})|^2]$ for all $k=1,\dots,N$.
To tackle this multiobjective problem, following the standard scalarization approach \cite{Rose}, here we propose to design quantizers that minimize the scalarized weighted mean squared error
\be \label{eq:wmse}
\sum_{k=1}^{N}\alpha_{k}E\left[\left|\mathbf{h}_{k}^{H}(\mathbf{w}_{\mathbf{H},k}s_{k}-\hat{\mathbf{x}})\right|^2\right],
\ee
where the weights $\alpha_{k}$ can be selected to enforce some fairness criterion (see \cite{Miettinen}).
We emphasize that the expectation in (\ref{eq:wmse}) is taken with respect also to the channel, hence making the codebooks $\hat{\mathcal{X}}_{i}$ for $i=1,\dots,M$, obtained through the minimization of (\ref{eq:wmse}), dependent only on long-term CSI, as mentioned above.
We also note that (\ref{eq:wmse}) differs from standard quantization error metrics, such as the error vector magnitude (EVM) or mean squared error \cite{Samardzija}-\cite{Si},
since (\ref{eq:wmse}) directly captures the overall system performance while the mentioned metrics apply on a per-fronthaul link basis (see also Sec. \ref{ch:SQ}).

\begin{figure}[t]
   \centering
    \begin{tabular}{cc}
    \begin{subfigure}{.5\textwidth}
    \includegraphics[scale=0.4]{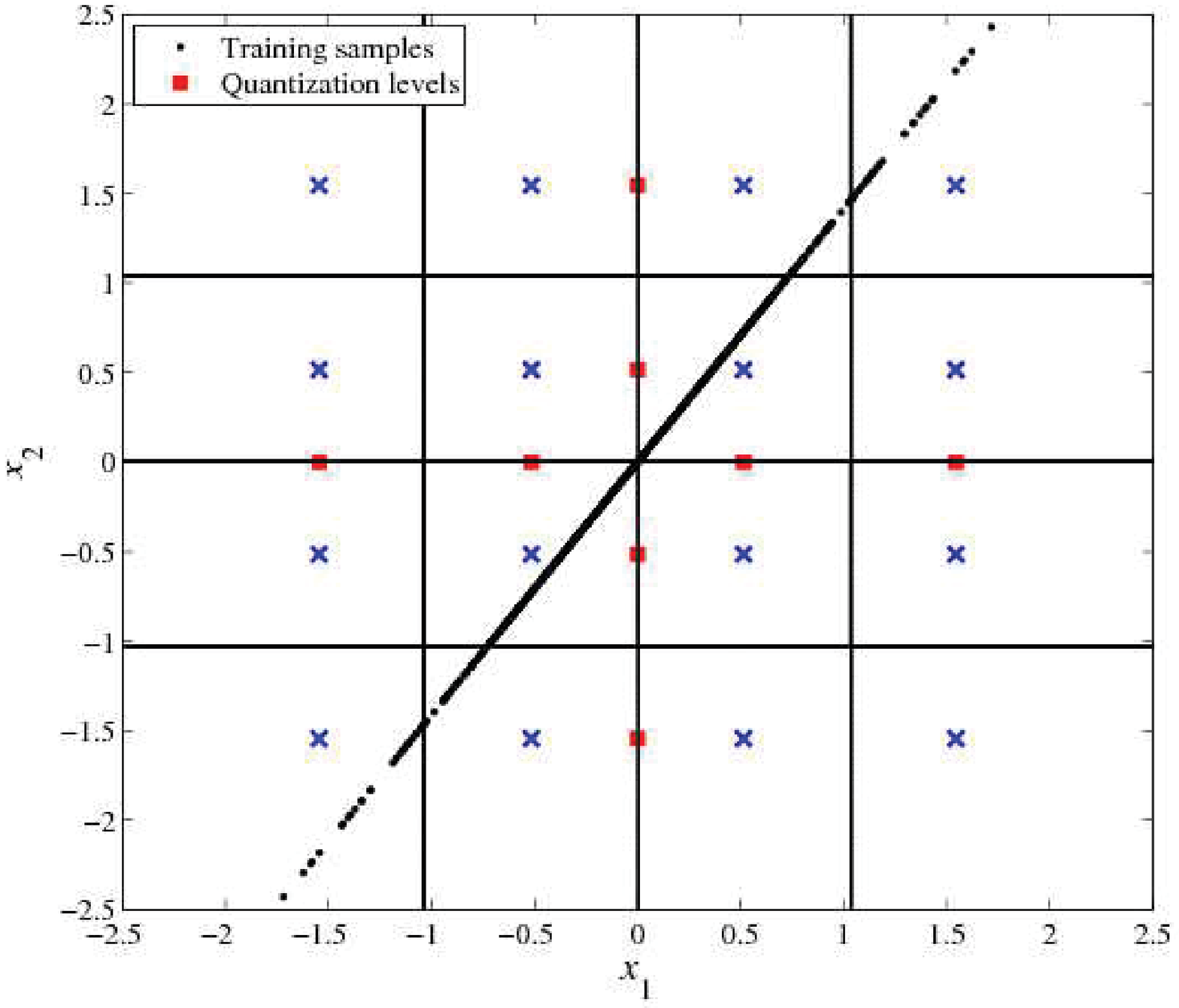}
    \caption{}
    \end{subfigure}
    \begin{subfigure}{.5\textwidth}
    \includegraphics[scale=0.4]{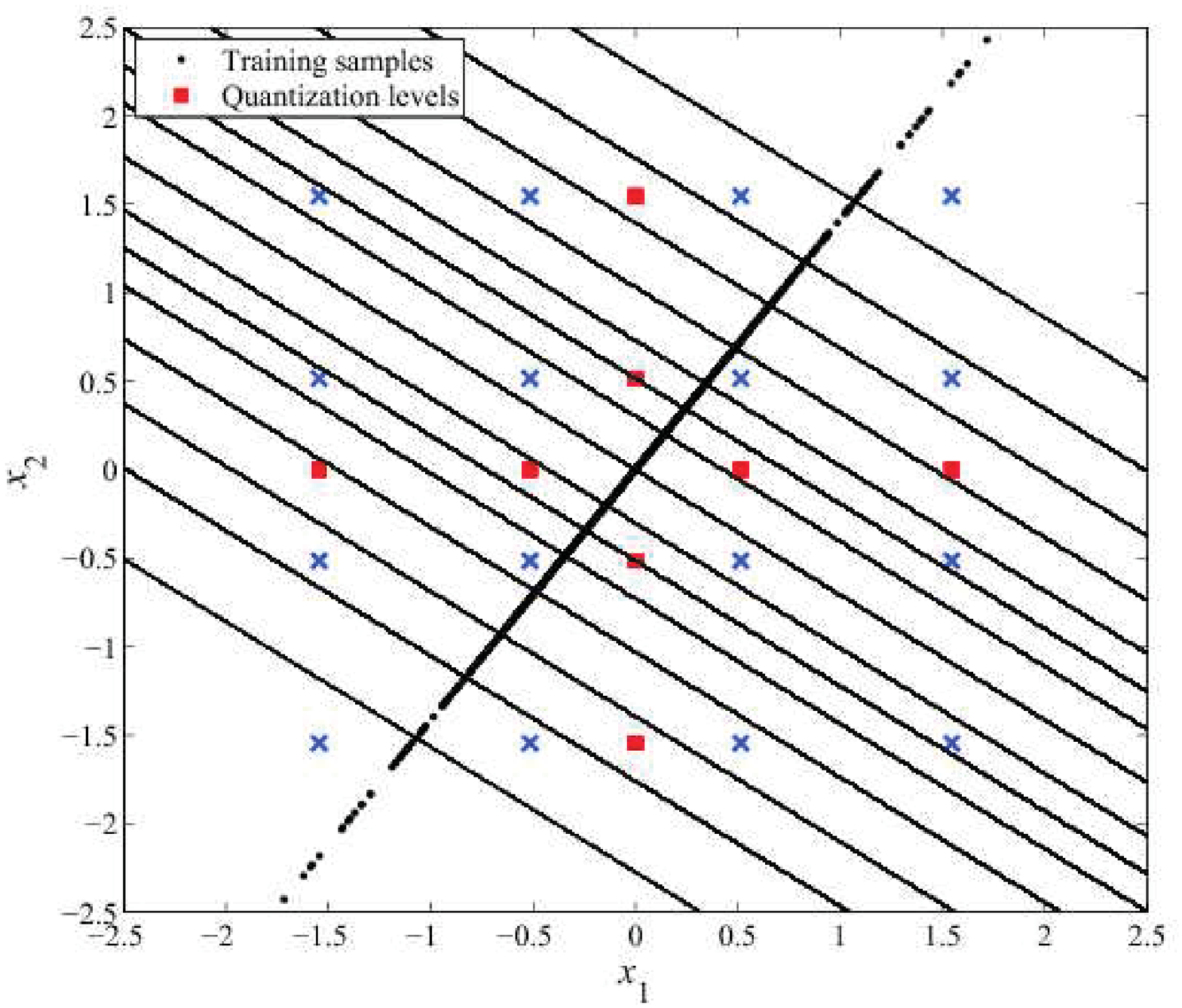}
    \caption{}
    \end{subfigure}
    \end{tabular}
   \caption{Illustration of: (a) conventional Point-to-Point Quantization (PtPQ); (b) Multivariate Quantization (MQ).}\label{fig:cb}
\end{figure}

\section{Introduction to Multivariate Quantization} \label{ch:MQ_principle}
In this section, we present intuitive arguments to illustrate the basic principles and potential benefits of MQ.
This is done by contrasting MQ with standard Point-to-Point Quantization (PtPQ) that operates separately on each fronthaul link.
To this end, we focus on the case of a single UE, i.e., $N=1$ and two RU, i.e., $M=2$, and assume for simplicity of visualization a real-valued system model.
Moreover, to further simplify the discussion, we adopt the matched beamformer $\mathbf{w}_{\mathbf{H}}=\mathbf{h}$, where $\mathbf{h}$ is the $2\times 1$ (real) channel vector for the given UE, and we have dropped the UE subscript to simplify the notation.
The transmitted signal (\ref{eq:precode_x}) can hence be written as $\mathbf{x}=\mathbf{h}s$, and some realization of $\mathbf{x}$ are shown as dots along the $55$ degree line in Fig. \ref{fig:cb} under the assumption that $s$ is a Gaussian random variable and the channel vector is $\mathbf{h}=[\sqrt{1/3},\sqrt{2/3}]^{T}$.
The figure also shows as squares on the horizontal and vertical axes the quantization levels that define the quantization codebooks $\hat{\mathcal{X}}_{1}$ and $\hat{\mathcal{X}}_{2}$ for the two RUs.
The two codebooks yield quantization points on the plane given by the indicated cross markers.
Note that the codebooks are the same for both conventional PtPQ (Fig. \ref{fig:cb}(a)) and MQ (Fig. \ref{fig:cb}(b)), as further discussed below.
As mentioned, each RU is informed only about its own codebook.

Because of quantization, the signal $\hat{\mathbf{x}}$ sent by two RUs must correspond to one of the quantization points (crosses) on the plane.
Therefore, the quantization error $(\mathbf{h}s-\hat{\mathbf{x}})$ between the desired signal, which lies on the $55^{\circ}$ line, and the selected point (cross) should be considered as a disturbance to the reception of the UE.
The key observation is that the impact of the quantization error $(\mathbf{h}s-\hat{\mathbf{x}})$ on the reception of the UE depends, by (\ref{eq:rec_Sig})-(\ref{eq:effSNR}), solely on the power $|\mathbf{h}^{T}(\mathbf{h}s-\hat{\mathbf{x}})|^{2}$ of its projection on the $55^{\circ}$ line.
That is, the only component of the quantization error that affects the UE is its projection onto the signal subspace.

Fig. \ref{fig:cb}(a) illustrates the quantization regions resulting from standard uniform PtPQ of the signals to be transmitted by the two RUs.
As seen in the figure, with PtPQ, the shape of the quantization regions is constrained to be rectangular.
Therefore, it is not possible for the CU to shape the quantization regions as a function of current CSI so as to minimize the projection of the quantization error onto the subspace occupied by the signal.

The limitation identified above can be alleviated by MQ, whereby quantization is performed jointly for the two RUs, as illustrated in Fig. \ref{fig:cb}(b).
We emphasize again that the same quantization codebooks are used as for PtPQ so that the RUs need not be informed about any change in the operation at the CU as a function of current CSI.
An optimized shape of the quantization region with MQ, using algorithms to be discussed in the rest of this paper, is illustrated in Fig. \ref{fig:cb}(b).
As seen in the figure, MQ enables the shaping of the quantization regions, with the aim of allowing a finer control of the impact of the quantization error on the received signals.
In the example of Fig. \ref{fig:cb}(a), in particular, the quantization error is such that the projection onto the subspace occupied by the signal is minimized.
MQ hence plays a complementary role to precoding: while precoding decides which ``spatial directions'' should be occupied by the signal, MQ controls which ``spatial directions'' are mostly affected by the quantization error.

\section{Multivariate Quantization with Fixed Precoding and Codebooks} \label{ch:MQ_fPC}
In this section, we study the baseline case in which the precoding matrix $\mathbf{W}_{\mathbf{H}}$ is fixed and the quantization codebooks are also given.
The quantization codebooks for RU $i$ is defined as $\hat{\mathcal{X}}_{i}=\{\hat{x}_{i}^{(1)},\dots,\hat{x}_{i}^{(2^{B_{i}})}\}$, which includes the $2^{B_{i}}$ possible quantization levels $\hat{x}_{i}^{(j)}$ for $j = 1,\dots,2^{B_{i}}$,
while the codebooks can be optimized based on long-term CSI.

\subsection{Conventional Point-to-Point Quantization (PtPQ)}
With conventional PtPQ, the CU independently carries out the quantization of the precoded signal $x_{1},\dots,x_{M}$ for each RUs.
For each RU $i$, the CU selects the quantization level $\hat{x}_{i}^{(f_{\hat{\mathcal{X}}_{i}}(x_{i}))}$, where
\be \label{eq:g_SQ}
f_{\hat{\mathcal{X}}_{i}}(x_{i})= \arg\min_{j}\left(x_{i}-\hat{x}_{i}^{(j)}\right)^2
\ee
is the standard minimum distance quantization function, which ties are broken arbitrarily.

\subsection{Multivariate Quantization (MQ)} \label{ch:sub_MQ_fPC}
With MQ, following the discussion in the previous section, the CU maps the precoded signals $\mathbf{w}_{\mathbf{H},k}s_{k}$ for all $k=1,\dots,N$ jointly across all RUs to the signal $\hat{\mathbf{x}}$ to be transmitted.
As discussed, a quantizer consists of two elements, namely a quantization codebook and a mapping \cite{Gray}.
For MQ, the quantization codebook $\hat{\mathcal{X}}=\hat{\mathcal{X}}_{1}\times\cdots\times \hat{\mathcal{X}}_{M}$ is given by the Cartesian product of the sets $\hat{\mathcal{X}}_{i}=\{\hat{x}_{i}^{(1)},\dots,\hat{x}_{i}^{(2^{B_{i}})}\}$ of the quantization levels $\hat{x}_{i}^{(j)}$ for each RU $i$ (i.e., the crosses in Fig. \ref{fig:cb}(b)).
The mapping instead is a function $f_{\hat{\mathcal{X}},\mathbf{H}}(\mathbf{x})$ that takes as input the baseband signal $\mathbf{x}$ in (\ref{eq:precode_x}), the CSI $\mathbf{H}$, and the codebook $\hat{\mathcal{X}}$, and outputs the corresponding quantization levels $[\hat{x}_{1}^{(j_{1})},\dots,\hat{x}_{M}^{(j_{M})}]^{T}$ or, equivalently, their indices $[j_{1},\dots,j_{M}]$.
The mapping defines the quantization regions illustrated in the example of Fig. \ref{fig:cb}.

The optimal mapping, from (\ref{eq:wmse}), is given by the function
\be \label{eq:g_MQ}
f_{\hat{\mathcal{X}},\mathbf{H}}(\mathbf{x}) = \arg\min_{j_{1},\dots,j_{M}} \sum_{k=1}^{N}\alpha_{k} \left|\mathbf{h}_{k}^{H}\left(\mathbf{w}_{\mathbf{H},k}s_{k}-\hat{\mathbf{x}}^{(j_{1},\dots,j_{M})}\right)\right|^{2},
\ee
where $\hat{\mathbf{x}}^{(j_{1},\dots,j_{M})}=[\hat{x}_{1}^{(j_{1})},\dots,\hat{x}_{M}^{(j_{M})}]$ is such that $x_{i}^{(j_{i})}\in\hat{\mathcal{X}}_{i}$, and we have made explicit the dependence of the function on both the codebook $\mathcal{\hat{X}}$ and on the channel $\mathbf{H}$.

\begin{figure}[t]
   \centering
    \begin{tabular}{cc}
    \begin{subfigure}{.4\textwidth}
    \includegraphics[scale=0.3]{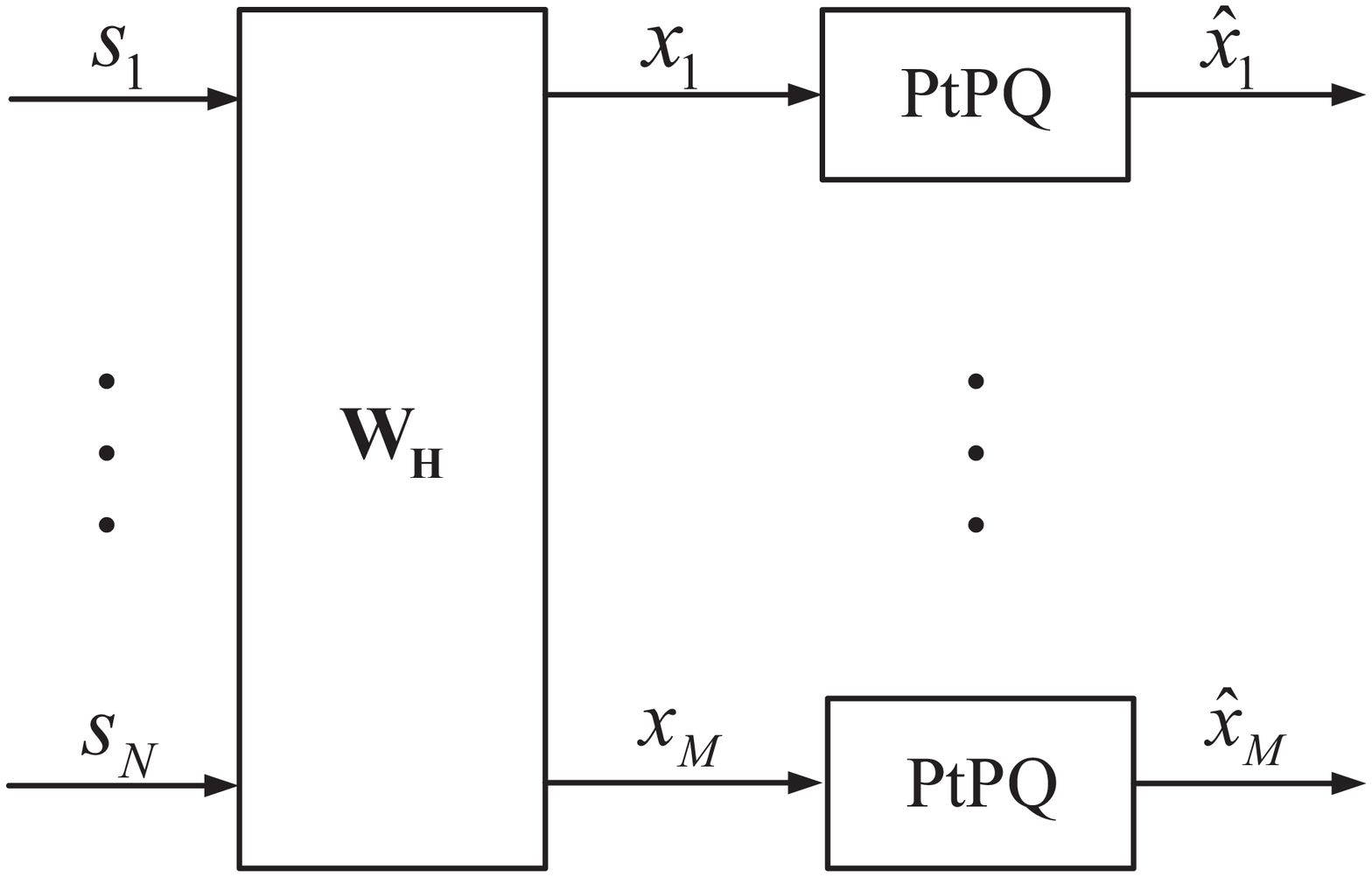}
    \caption{}
    \end{subfigure}
    \begin{subfigure}{.4\textwidth}
    \includegraphics[scale=0.3]{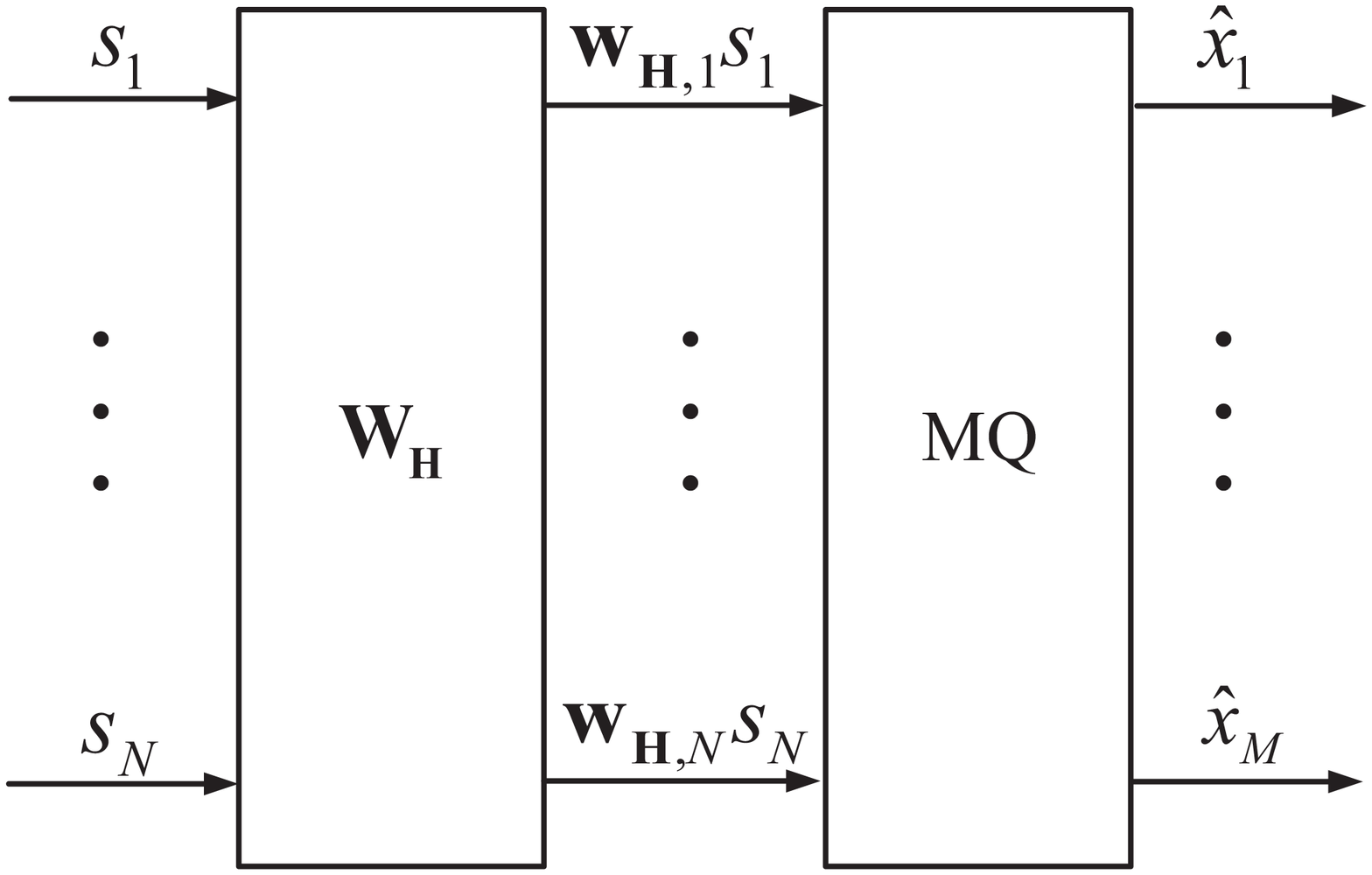}
    \caption{}
    \end{subfigure}
    \end{tabular}
   \caption{Block diagram of: (a) conventional Point-to-Point Quantization (PtPQ); (b) Multivariate Quantization (MQ).}\label{fig:block}
\end{figure}

\section{Multivariate Quantization Design with Fixed Precoding} \label{ch:MQ_sep}
In this section, we will first review the codebook design of conventional PtPQ in Sec. \ref{ch:SQ} and then introduce the proposed MQ codebook design in Sec. \ref{ch:MQ}.
Throughout this section, we assume that the downlink precoder $\mathbf{W}_{\mathbf{H}}$ is fixed and not subject to optimization.

\begin{algorithm}[t]
  \caption{Point-to-Point Quantization (PtPQ) Design}
  \begin{algorithmic}
    \State \textbf{Input}: Generate $N_{s}$ training values $\mathcal{S}=\{\mathbf{s}(1),\dots,\mathbf{s}(N_{s})\}$. Set a threshold $\epsilon \geq 0$.
    \State \textbf{Initialization}: Initialize the codebook as $\hat{\mathcal{X}}_{i}^{[0]}=\{\hat{x}_{i}^{(j)}:j=1,\dots,2^{B_{i}}\}$ and set $t=0$ and $D_{-1}=\infty$.
    \Repeat
    \State \parbox[t]{\dimexpr\linewidth-\algorithmicindent}{$\bullet$ Find the minimum-distortion partition of $\mathcal{S}$ as $\mathcal{S}_{i}^{(j)}=\{\mathbf{s}\in \mathcal{S}:f_{\hat{\mathcal{X}}_{i}^{[t]}}(\mathbf{w}_{\mathbf{H},i}\mathbf{s})=j\}$, where function $f_{\hat{\mathcal{X}}_{i}^{[t]}}$ is defined in (\ref{eq:g_SQ}) with $j=1,\dots,2^{B_{i}}$.}
    \State $\bullet$ Compute the average distortion $D^{[t]}=1/N_{s}\sum_{n=1}^{N_{s}}(\mathbf{w}_{\mathbf{H},i}\mathbf{s}(n)-\hat{x}_{i}^{(f_{\hat{\mathcal{X}}_{i}^{[t]}}(\mathbf{w}_{\mathbf{H},i}\mathbf{s}(n)))})^2$.
    \State $\bullet$ Obtain the new codebook as $\hat{\mathcal{X}}_{i}^{[t+1]}=\{\hat{x}_{i}^{(j)}:j=1,\dots,2^{B_{i}}\}$ by solving the quadratic convex problem (\ref{eq:Up_cb_SQ}).
    \State $\bullet$ Set $t=t+1$.
    \Until {$(D^{[t-1]}-D^{[t]})/D^{[t]}\leq \epsilon$}.
    \State \textbf{Output}: Codebook $\hat{\mathcal{X}}_{i}=\hat{\mathcal{X}}_{i}^{[t]}$.
  \end{algorithmic}
\label{al:SQ}
\end{algorithm}

\subsection{Conventional Point-to-Point Quantization} \label{ch:SQ}
With PtPQ, the CU designs the quantization function $Q_{i}(x_{i})$ for each RU $i$, where $Q_{i}(\cdot)$ takes values in a codebook $\hat{\mathcal{X}}_{i}$.
Following the discussion in the previous section, we observe that PtPQ cannot directly tackle the minimization of (\ref{eq:wmse}) since the latter depends on the entire signal $\mathbf{x}$, while PtPQ prescribes that each quantized symbol $\hat{x}_{i}$ is a function of only $x_{i}$.
In other words, while MQ can adapt to the current realization of the channel $\bf{H}$ by choosing the quantized signal $\hat{\bf{x}}$ as a function of the entire vector $\bf{x}$, PtPQ is limited to operating on each RU separately and hence can only optimize a per-RU criterion.
Accordingly, we adopt the conventional design of each codebook $\hat{\mathcal{X}}_{i}$ that aims at the minimization of the EVM or mean squared error (see, e.g., \cite{Gray}), that is
\be \label{eq:opt_SQ}
\min_{\hat{\mathcal{X}}_{i}=\{\hat{x}_{i}^{(j)}\}_{j=1}^{2^{B_{i}}}} E\left[\left(x_{i}-\hat{x}_{i}^{(f_{\hat{\mathcal{X}}_{i}}(x_{i}))}\right)^2\right],
\ee
where the expectation is with respect to the transmitted symbols $\mathbf{s}$, and $f_{\hat{\mathcal{X}}_{i}}(\cdot)$ represents the optimal quantization mapping (\ref{eq:g_SQ}).
When designing the codebook using (\ref{eq:opt_SQ}), one needs to account for the power constraint (\ref{eq:per_power_cons}).
Problem (\ref{eq:opt_SQ}) can then be tackled by means of a simple extension of the training-based Lloyd-Max algorithm \cite{Gray} that accounts for the per-RU power constraint and is summarized in Algorithm \ref{al:SQ} for reference.
In the algorithm, the codebook $\hat{\mathcal{X}}_{i}=\{\hat{x}_{i}^{(j)}:j=1,\dots,2^{B_{i}}\}$ for RU $i$ is updated by solving the following convex quadratic problems
\be \label{eq:Up_cb_SQ}
\hat{\mathcal{X}}_{i} = \arg\min_{\{\hat{x}_{i}^{(j)}\}_{j=1}^{2^{B_{i}}}}&&\sum_{j=1}^{2^{B_{i}}}\sum_{n:~\mathbf{s}(n)\in \mathcal{S}_{i}^{(j)}}\left|\mathbf{w}_{\mathbf{H},i}\mathbf{s}(n)-\hat{x}_{i}^{(j)}\right|^{2} \nonumber \\
\textrm{s.t.} &&\sum_{j=1}^{2^{B_{i}}}p(\hat{x}_{i}^{(j)})|\hat{x}_{i}^{(j)}|^{2} \leq 1,
\ee
where $p(\hat{x}_{i}^{(j)})=|\mathcal{S}_{i}^{(j)}|/N_{s}$ is the fraction of the $N_{s}$ training samples that are quantized to $\hat{x}_{i}^{(j)}$ for RU $i$.
We finally observe that more advanced algorithms, such as deterministic annealing \cite{Rose}, could be used in lieu of Lloyd-Max, although this is not further elaborated here.

\subsection{Multivariate Quantization} \label{ch:MQ}
In this section, we detail a procedure for the optimization of codebook in MQ with the aim of minimizing the distortion criterion (\ref{eq:wmse}).
The design of the quantization codebook $\hat{\mathcal{X}}$ can be formulated as the problem
\be \label{eq:Opt_codebook}
\hat{\mathcal{X}}=\arg\min_{\{\hat{x}_{i}^{(j)}\}_{i=1}^{M}{}_{j=1}^{2^{B_{i}}}} &&\sum_{k=1}^{N}\alpha_{k}E\left[\left|\mathbf{h}_{k}^{H}\left(\mathbf{w}_{\mathbf{H},k}s_{k}-\hat{\mathbf{x}}^{(f_{\hat{\mathcal{X}},\mathbf{H}}(\mathbf{x}))}\right)\right|^2\right] \nonumber \\
\mathrm{s.t.} && E\left[\left|\hat{x}_{i}\right|^2\right] \leq 1~\textrm{for}~i=1,\dots,M,
\ee
where the expectations are taken with respect to $\mathbf{s}$ and $\mathbf{H}$.
We observe that, as per (\ref{eq:Opt_codebook}), the codebook $\hat{\mathcal{X}}$ is not a function of the instantaneous CSI but only of the distribution of $\mathbf{H}$.
This guarantees that the RUs need not be informed about a new codebook any time the channel $\mathbf{H}$ changes but only at the time scale of the variations of long-term CSI.
The RUs need also not be informed about the mapping (\ref{eq:g_MQ}).

\begin{algorithm} [t]
  \caption{Multivariate Quantization (MQ) Design}
  \begin{algorithmic}
    \State \textbf{Input}: Generate $N_{h}$ independent training channels $\mathcal{H}=\{\mathbf{H}(1),\dots,\mathbf{H}({N_{h}})\}$ from the known channel distribution and $N_{s}$ independent training values $\mathcal{S}=\{\mathbf{s}(1),\dots,\mathbf{s}(N_{s})\}$ from the known distribution of $\mathbf{s}$. Set a threshold $\epsilon \geq 0$.
    \State \textbf{Initialization}: Initialize the codebook as $\hat{\mathcal{X}}^{[1]}$ and set $t=1$ and $D^{[0]}=\infty$.
    \Repeat
    \For {$m=1$ to $N_{h}$}
        \State \parbox[t]{\dimexpr\linewidth-\algorithmicindent-\algorithmicindent\relax}{$\bullet$ Find the minimum-distortion partition of $\mathcal{S}$ when the channel is $\mathbf{H}(m)$ as $\mathcal{S}^{(j_{1},\dots,j_{M},m)}=\{\mathbf{s}\in \mathcal{S}: f_{\hat{\mathcal{X}}^{[t]},\mathbf{H}(m)}(\mathbf{W}_{\mathbf{H}(m)}\mathbf{s})=[j_{1},\dots,j_{M}]^T\}$ with $j_{i}=1,\dots,2^{B_{i}}$, $i=1,\dots,M$ and $m=1,\dots,N_{h}$.}
    \EndFor
    \State \parbox[t]{\dimexpr\linewidth-\algorithmicindent-\algorithmicindent\relax}{$\bullet$ \hspace{-6mm} Compute the average distortion $D^{[t]}=1/(N_{h}N_{s})\sum_{m=1}^{N_{h}}\sum_{n=1}^{N_{s}}\sum_{k=1}^{N}|\mathbf{h}_{k}(m)^{H}(\mathbf{w}_{\mathbf{H}(m),k}s_{k}(n)-\hat{\mathbf{x}}^{(f_{\hat{\mathcal{X}}^{[t]},\mathbf{H}(m)}(\mathbf{W}_{\mathbf{H}(m)}\mathbf{s}(n)))})|^2$.}
    \State \parbox[t]{\dimexpr\linewidth-\algorithmicindent}{$\bullet$ Obtain the updated codebook as $\hat{\mathcal{X}}^{[t+1]}$ by solving the quadratic convex problem (\ref{eq:Up_codebook}).}
    \State $\bullet$ Set $t=t+1$.
    \Until {$(D^{[t-1]}-D^{[t]})/D^{[t]}\leq \epsilon$}.
    \State \textbf{Output}: Codebook $\hat{\mathcal{X}}=\hat{\mathcal{X}}^{[t]}$.
  \end{algorithmic}
\label{al:MQ}
\end{algorithm}

\textit{\textbf{Codebook Optimization}}: In order to address the optimization (\ref{eq:Opt_codebook}) over the codebook $\hat{\mathcal{X}}$,
we follow the standard approach of the Lloyd-Max algorithm, and its extension to vector quantization due to Linde-Buzo-Gray \cite{LBG}, by iterating between the application of the mapping (\ref{eq:g_MQ}) for a fixed codebook and the optimization of the codebook (\ref{eq:Opt_codebook}) for the obtained mapping given the current codebook iterate.
The algorithm is based on randomly generated training samples for $\mathbf{s}$ and for $\mathbf{H}$, namely $\mathcal{S}=\{\mathbf{s}(1),\dots,\mathbf{s}(N_{s})\}$ and $\mathcal{H}=\{\mathbf{H}(1),\dots,\mathbf{H}({N_{h}})\}$, respectively, and is detailed in Algorithm \ref{al:MQ}.
We emphasize that the algorithm is run offline based only on long-term CSI.
Moreover, once the codebook $\hat{\mathcal{X}}$ is designed, the mapping (\ref{eq:g_MQ}) is applied for the given instantaneous CSI $\mathbf{H}$.

Referring to Algorithm \ref{al:MQ}, the proposed MQ design scheme updates the codebook $\hat{\mathcal{X}}$ by solving the quadratic convex problem
\be \label{eq:Up_codebook}
\hat{\mathcal{X}}^{[t+1]} = \arg\min_{\{\hat{x}_{i}^{(j_{i})}\}_{i=1}^{M}{}_{j_{i}=1}^{2^{B_{i}}}}&&\sum_{m=1}^{N_{h}}\sum_{j_{1},\cdots,j_{M}}\sum_{n:~\mathbf{s}(n)\in \mathcal{S}^{(j_{1},\dots,j_{M},m)}}\sum_{k=1}^{N}\alpha_{k}\left|\mathbf{h}_{k}(m)^{H}(\mathbf{w}_{\mathbf{H}(m),k}s_{k}(n)-\hat{\mathbf{x}}^{(j_{1},\dots,j_{M})})\right|^{2} \nonumber \\
\textrm{s.t.}&&\sum_{j_{i}=1}^{2^{B_{i}}}p(\hat{x}_{i}^{(j_{i})})|\hat{x}_{i}^{(j_{i})}|^{2} \leq 1 ~\textrm{for}~i=1,\dots,M,
\ee
with $\sum_{i=1}^{M}2^{B_{i}}$ unknown variables, where $p(\hat{x}_{i}^{(j_{i})})=\sum_{m=1}^{N_{h}}\sum_{j_{k}\neq j_{i}}\left|\mathcal{S}^{(j_{1},\dots,j_{M},m)}\right|\big/N_{h}N_{s}$ is the fraction of the $N_{h}N_{s}$ training samples that are quantized to $\hat{x}_{i}^{(j_{i})}$ for each RU $i$,
and the set $\mathcal{S}^{(j_{1},\dots,j_{M},m)}$ contains all training samples in $\mathcal{S}$ that are mapped to the quantization index $[j_{1},\dots,j_{M}]$ when the CSI is $\mathbf{H}(m)$ by mapping (\ref{eq:g_MQ}).
As a final note, we observe that, for ease of notation, the problem (\ref{eq:Up_codebook}) can be made more explicit by defining the $\sum_{i=1}^{M}2^{B_{i}} \times 1$ vector of unknown variables $\hat{\mathbf{x}}=[\hat{x}_{1}^{(1)},\dots,\hat{x}_{1}^{(2^{B_{1}})},\dots,\hat{x}_{M}^{(1)},\dots,\hat{x}_{M}^{(2^{B_{M}})}]^{T}$
and the $M \times \sum_{i=1}^{M}2^{B_{i}}$ transformation matrix $T^{(j_{1},\dots,j_{M})}$ such that $\hat{\mathbf{x}}^{(j_{1},\dots,j_{M})}=T^{(j_{1},\dots,j_{M})}\hat{\mathbf{x}}$ as
\be
\hat{\mathcal{X}}^{[t+1]} = \arg\min_{\{\hat{x}_{i}^{(j_{i})}\}_{i=1}^{M}{}_{j_{i}=1}^{2^{B_{i}}}} \hat{\mathbf{x}}^{H}\left\{\sum_{m=1}^{N_{h}}\sum_{j_{1},\cdots,j_{M}}\left|\mathcal{S}^{(j_{1},\dots,j_{M},m)}\right|\left(T^{(j_{1},\dots,j_{M},m)}\right)^{H}\sum_{k=1}^{N}\alpha_{k}\mathbf{h}_{k}(m)\mathbf{h}_{k}(m)^{H}T^{(j_{1},\dots,j_{M},m)}\right\}\hat{\mathbf{x}}
\ee
under the same constraint.
We finally observe that the complexity of the MQ codebook design in Algorithm \ref{al:MQ} in each iteration is given by the sum of the complexities of computing the minimum distortion partition of the training set $\mathcal{S}$ and of solving the quadratic convex problem (\ref{eq:Up_codebook}).
The first is given by $N_{h}N_{s}2^{\sum_{i=1}^{M}B_{i}}$ comparisons, while the second is polynomial in the size of the problem, which includes $M2^{\sum_{i=1}^{M}B_{i}}$ complex numbers.
While this complexity may be high if $M$ is large, we note that this optimization can be carried out offline.
Furthermore, as we will see in Sec. \ref{ch:NR}, MQ performs well even with fixed, non-optimized, codebooks.
Finally, additional discussion in complexity can be found in Sec. \ref{ch:sbMQ}.

\section{Joint Design of Multivariate Quantization and Precoding} \label{ch:MQ_JD}
In the previous section, we have proposed a method that designs MQ for a fixed precoding matrix $\mathbf{W}_{\mathbf{H}}$, which is generally selected as a function of the channel realization $\mathbf{H}$.
As discussed in Sec. \ref{ch:MQ_principle}, however, there is an important interplay between precoding and MQ, since precoding determines the spatial dimensions occupied by the signal, while MQ shapes the spatial correlation of the disturbance due to quantization.
Based on this observation, in this section, we aim at jointly optimizing the precoding matrix $\mathbf{W}_{\mathbf{H}}$ and MQ.
This was done in \cite{Park:TSP} under the assumption of information-theoretically optimal block coding, while here we consider symbol-by-symbol quantization.

As in the previous section, the first step is to optimize offline the MQ codebook $\hat{\mathcal{X}}$ based on long-term CSI.
To this end, the proposed approach is based on a heuristic iterative procedure that alternates between the optimization of matrices $\mathbf{W}_{\mathbf{H}}$ and of the MQ codebook.
Specifically, since each precoding matrix $\mathbf{W}_{\mathbf{H}}$ affects both numerator and denominator in the effective SNR (\ref{eq:effSNR}), when optimizing over $\mathbf{W}_{\mathbf{H}}$,
we aim at maximizing a measure of the sum-rate $\sum_{k=1}^{N}\log_{2}(1+\textrm{SNR}_{k}^{\textrm{eff}})$ for fixed MQ codebook.
Instead, when optimizing over the MQ codebook, we fix matrices $\mathbf{W}_{\mathbf{H}}$, and proceed as in the previous section.
Details of the algorithm summarized in Algorithm \ref{al:MQ_MM} are provided in the rest of this section.

We now discuss the optimization of the precoding matrix $\mathbf{W}_{\mathbf{H}}$ for fixed MQ.
The direct optimization of the achievable rate appears challenging due to the non-linear dependence of the effective SNR in (\ref{eq:effSNR}) on $\mathbf{W}_{\mathbf{H}}$ via the quantized signal $\hat{\mathbf{x}}$.
To address this problem, we approximate the effect of MQ by means of an additive quantization noise model.
In particular, we model the transmitted signal as
\be \label{eq:q_x_saqnm}
\hat{\mathbf{x}}=\mathbf{W}_{\mathbf{H}}\mathbf{s}+\mathbf{q},
\ee
where the $M \times 1$ quantization noise vector $\mathbf{q}$ is assumed to follow a complex Gaussian distribution $\mathcal{CN}(\mathbf{0},\mathbf{\Omega}_{\mathbf{H}})$, with a channel-dependent covariance matrix $\mathbf{\Omega}_{\mathbf{H}}$, and to be independent of $\mathbf{s}$.
The covariance matrix $\mathbf{\Omega}_{\mathbf{H}}$ accounts for the correlation that is made possible by the use of MQ (recall Fig. \ref{fig:cb}) and is estimated by using training samples for the given MQ codebook as discussed below.
Using (\ref{eq:q_x_saqnm}), the achievable rate for the UE $k$ can be computed as (see, e.g., \cite{Park:TSP})
\be \label{eq:Rk}
R_{k} &=& \log\det\left(\mathbf{I}+{P}\mathbf{h}_{k}^{H}\left(\mathbf{W}_{\mathbf{H}}\mathbf{W}_{\mathbf{H}}^{H}+\mathbf{\Omega}_{\mathbf{H}}\right)\mathbf{h}_{k}\right)-\log\det\left(\mathbf{I}+{P}\mathbf{h}_{k}^{H}\left(\sum_{l\neq k}{\mathbf{w}}_{{\mathbf{H}},l}{\mathbf{w}}_{{\mathbf{H}},l}^{H}+\mathbf{\Omega}_{\mathbf{H}}\right)\mathbf{h}_{k}\right),
\ee
where we recall that ${\mathbf{w}}_{{\mathbf{H}},l}$ is the $M \times 1$ column vector corresponding to the UE $l$ from the precoding matrix $\mathbf{W}_{{\mathbf{H}}}$.
We propose to maximize the sum-rate $\sum_{k}R_{k}$ with (\ref{eq:Rk}) under the per-RU power constraints (\ref{eq:per_power_cons}), which, under model (\ref{eq:q_x_saqnm}), can be written as
\be
\|\mathbf{e}_{i}^{T}\mathbf{W}_{{\mathbf{H}}}\|^{2}+\mathbf{e}_{i}^{T}\mathbf{\Omega}_{\mathbf{H}}\mathbf{e}_{i} \leq 1,
\ee
for $i=1,\dots,M$, where the $M \times 1$ vector $\mathbf{e}_{i}$ has all zero elements except for the $i$th element of which is set to $1$.

\begin{algorithm} [t]
  \caption{Joint Design of Precoding matrix and MQ}
  \begin{algorithmic}
    \State \textbf{Input}: Generate $N_{h}$ independent training channels $\mathcal{H}=\{\mathbf{H}(1),\dots,\mathbf{H}({N_{h}})\}$ from the known channel distribution and $N_{s}$ independent training values $\mathcal{S}=\{\mathbf{s}(1),\dots,\mathbf{s}(N_{s})\}$ from the known distribution of $\mathbf{s}$. Set a threshold $\epsilon \geq 0$.
    \State \textbf{Initialization}: Initialize the codebook as $\hat{\mathcal{X}}^{[1]}$ and quantization error covariance matrices to $\mathbf{\Omega}_{\mathbf{H}(m)}^{[1]}=\mathbf{0}$ for all $m=1,\dots,N_{h}$. Set $t=1$ and $D^{[0]}=\infty$.
    \Repeat
    \For {$m=1$ to $N_{h}$}
    \State \parbox[t]{\dimexpr\linewidth-\algorithmicindent-\algorithmicindent\relax}{$\bullet$ Perform precoding optimization via the DC algorithm (Algorithm \ref{al:DC}) with input $\mathbf{H}(m)$ and $\mathbf{\Omega}_{\mathbf{H}(m)}^{[t]}$, and output $\mathbf{W}_{\mathbf{H}(m)}^{[t]}$.}
    \State \parbox[t]{\dimexpr\linewidth-\algorithmicindent-\algorithmicindent\relax}{$\bullet$ $\hspace{-3mm}$ Obtain the minimum-distortion partition of $\mathcal{S}$ when the channel is $\mathbf{H}(m)$ as $\mathcal{S}^{(j_{1},\dots,j_{M},m)}=\{\mathbf{s}\in \mathcal{S}: f_{\hat{\mathcal{X}}^{[t]},\mathbf{H}(m)}(\mathbf{W}_{\mathbf{H}(m)}^{[t]}\mathbf{s})=[j_{1},\dots,j_{M}]^T\}$ with $j_{i}=1,\dots,2^{B_{i}}$, $i=1,\dots,M$.}
    \EndFor
    \State \parbox[t]{\dimexpr\linewidth-\algorithmicindent}{$\bullet$ Update the codebook as $\hat{\mathcal{X}}^{[t+1]}$ by solving the optimization problem (\ref{eq:Up_codebook}).}
    \State \parbox[t]{\dimexpr\linewidth-\algorithmicindent-\algorithmicindent\relax}{$\bullet$ $\hspace{-5mm}$ Compute the average distortion $D^{[t]}=1/(N_{h}N_{s})\sum_{m=1}^{N_{h}}\sum_{n=1}^{N_{s}}\sum_{k=1}^{N}$ $|\mathbf{h}_{k}(m)^{H}(\mathbf{w}_{\mathbf{H}(m),k}^{[t]}s_{k}(n)-\hat{\mathbf{x}}^{(f_{\hat{\mathcal{X}}^{[t]},\mathbf{H}(m)}(\mathbf{W}_{\mathbf{H}(m)}^{[t]}\mathbf{s}(n)))})|^2$.}
    \State \parbox[t]{\dimexpr\linewidth-\algorithmicindent}{$\bullet$ Estimate the covariance matrix $\mathbf{\Omega}_{\mathbf{H}(m)}^{[t]}$ by using (\ref{eq:Up_Omega}) for every $m=1,\dots,N_{h}$.}
    \State $\bullet$ Set $t=t+1$.
    \Until {$(D^{[t-1]}-D^{[t]})/D^{[t]}\leq \epsilon$}.
    \State \textbf{Output}: Codebook $\hat{\mathcal{X}}=\hat{\mathcal{X}}^{[t]}$.
  \end{algorithmic}
\label{al:MQ_MM}
\end{algorithm}
\begin{algorithm} [t!]
  \caption{Difference-of-Convex (DC) Algorithm for Precoding Optimization}
  \begin{algorithmic}
    \State \textbf{Input}: Channel $\mathbf{H}$, covariance matrix $\mathbf{\Omega}_{\mathbf{H}}$, and maximum iteration number $r_{\max}$.
    \State \parbox[t]{\dimexpr\linewidth-\algorithmicindent}{\textbf{Initialization}: Initialize $\mathbf{W}_{\mathbf{H}}^{(1)}=\{\mathbf{w}_{\mathbf{H},1}^{(1)},\dots,\mathbf{w}_{\mathbf{H},N}^{(1)}\}$ and set $\mathbf{V}_{\mathbf{H},k}^{(1)}=\mathbf{w}_{\mathbf{H},k}^{(1)}(\mathbf{w}_{\mathbf{H},k}^{(1)})^{H}$ for $k=1,\dots,N$ and $r=1$.}
    \Repeat
    \State $\bullet$ $\mathbf{V}_{\mathbf{H}}^{(r+1)}$ = $\arg\max\limits_{\mathbf{V}_{\mathbf{H}}}\sum_{k=1}^{N}\tilde{R}_{k}(\mathbf{V}_{\mathbf{H}}|\mathbf{V}_{\mathbf{H}}^{(r)},\mathbf{\Omega}_{\mathbf{H}})$ in (\ref{eq:Rk_MM})
    \State \parbox[t]{\dimexpr\linewidth-\algorithmicindent}{$\hspace{24mm} \mathrm{s.t.}~\mathbf{e}_{i}^{T}(\sum_{k=1}^{N}\mathbf{V}_{\mathbf{H},k}+\mathbf{\Omega}_{\mathbf{H}})\mathbf{e}_{i} \leq 1$ for $i=1,\dots,M$}
    \State $\hspace{30mm} \mathbf{V}_{\mathbf{H},k}\succeq \mathbf{0}$ for $k=1,\dots,N$.
    \State $\bullet$ Set $r=r+1$.
    \Until $r=r_{\max}$.
    \State \textbf{Output}: Precoding matrix $\mathbf{W}_{\mathbf{H}}=[\mathbf{w}_{\mathbf{H},1}^{(r_{\max})},\dots,\mathbf{w}_{\mathbf{H},N}^{(r_{\max})}]$, where $\mathbf{w}_{\mathbf{H},k}^{(r_{\max})}$ is obtained from the covariance matrix $\mathbf{V}_{\mathbf{H},k}^{(r_{\max})}$ via rank-$1$ reduction.
  \end{algorithmic}
\label{al:DC}
\end{algorithm}

The design problem for the precoding matrix $\mathbf{W}_{\mathbf{H}}$ outlined above is tackled by first making the change of variables $\mathbf{V}_{\mathbf{H},k}={\mathbf{w}}_{\mathbf{H},k}{\mathbf{w}}_{\mathbf{H},k}^{H}$ and dropping the rank-$1$ constraint on matrices $\mathbf{V}_{\mathbf{H},k}$ for $k=1,\dots,N$.
The resulting problem can be solved to a local minimum by following the Difference-of-Convex (DC) algorithm \cite{Tao}, as discussed in e.g., \cite{Park:TSP}.
The detailed procedure for jointly optimizing the precoding matrix is summarized in Algorithm \ref{al:DC}.
In this algorithm, we have defined the concave upper bound on (\ref{eq:Rk}) obtained by linearizing the second term in (\ref{eq:Rk}) around the current iterate $\mathbf{V}_{\mathbf{H}}^{[r]}$ as
\be \label{eq:Rk_MM}
\tilde{R}_{k}\left(\mathbf{V_{\mathbf{H}}}|\mathbf{V}_{\mathbf{H}}^{[r]},\mathbf{\Omega}_{\mathbf{H}}\right) &\triangleq& \log\det\left(\mathbf{I}+{P}\mathbf{h}_{k}^{H}\left(\sum_{l=1}^{N}\mathbf{V}_{\mathbf{H},l}+\mathbf{\Omega}_{\mathbf{H}}\right)\mathbf{h}_{k}\right)\nonumber \\
&& \hspace{10mm} -f\left(\mathbf{I}+{P}\mathbf{h}_{k}^{H}\left(\sum_{l\neq k}\mathbf{V}_{\mathbf{H},l}^{[r]}+\mathbf{\Omega}_{\mathbf{H}}\right)\mathbf{h}_{k},\mathbf{I}+{P}\mathbf{h}_{k}^{H}\left(\sum_{l\neq k}\mathbf{V}_{\mathbf{H},l}+\mathbf{\Omega}_{\mathbf{H}}\right)\mathbf{h}_{k}\right),
\ee
where the function $f(\mathbf{A},\mathbf{B})$ is obtained from the first-order Taylor expansion of the log-det function
\be
f(\mathbf{A},\mathbf{B})=\log\det(\mathbf{A})+\frac{1}{\ln 2}\mathrm{tr}(\mathbf{A}^{-1}(\mathbf{B}-\mathbf{A})).
\ee
As seen in Algorithm \ref{al:DC}, the obtained solution $\mathbf{V}_{\mathbf{H}}^{*}$ is used to calculate the precoding matrix $\mathbf{W}_{\mathbf{H}}$ by using the standard rank-reduction approach.

The complete algorithm is shown in Algorithm \ref{al:MQ_MM}.
As discussed, it consists of an iterative execution of the DC algorithm introduced above for the optimization of precoding and of the procedure proposed in the previous section for the design of the MQ codebook.
The interface between the two optimizations is given by the estimate of the quantization noise covariance matrix, which is given by
\be \label{eq:Up_Omega}
\mathbf{\Omega}_{\mathbf{H}}=\frac{1}{N_{s}}\sum_{n=1}^{N_{s}}\left(\mathbf{W}_{\mathbf{H}}\mathbf{s}(n)-\hat{\mathbf{x}}^{(f_{\hat{\mathcal{X}},\mathbf{H}}(\mathbf{W}_{\mathbf{H}}\mathbf{s}(n)))}\right)\left(\mathbf{W}_{\mathbf{H}}\mathbf{s}(n)-\hat{\mathbf{x}}^{(f_{\hat{\mathcal{X}},\mathbf{H}}(\mathbf{W}_{\mathbf{H}}\mathbf{s}(n)))}\right)^{H},
\ee
based on the current iterates $\mathbf{W}_{\mathbf{H}}$ and $\hat{\mathcal{X}}$ for precoding and MQ codebook, respectively.

\section{Reduced-Complexity Multivariate Quantization via Successive Block Quantization} \label{ch:sbMQ}
MQ requires to perform a joint mapping, using function (\ref{eq:g_MQ}), between the input vector $\bf{x}$ and the output $\hat{\bf{x}}$. 
This entails a search over $2^{\sum_{i=1}^{M}B_{i}}$ possible values in the codebook $\hat{\mathcal{X}}=\hat{\mathcal{X}}_{1}\times\cdots\times\hat{\mathcal{X}}_{M}$, and hence, it may entail a high computational complexity as the number of RUs is large.
Note that, in contrast, PtPQ only requires a search over $2^{B_{i}}$ possible values for any RU $i$.
To tackle this issue, in this section, we propose a reduced-complexity implementation of MQ by successive block quantization steps.
To keep notation at a minimum, we focus on fixed codebook and precoding.

We first detail the proposed scheme for an implementation with blocks of size $d=1$.
The proposed successive MQ strategy entails $M$ sequential PtPQ step, whereby the CU quantizes the signal $x_{i}$ for RU $i$ to the codeword $\hat{x}_{i}\in\hat{\mathcal{X}}_{i}$ by using the previously quantized signals $\hat{x}_{1},\dots,\hat{x}_{i-1}$ as seen in Fig. \ref{fig:SMQ}.
Note that, since each precoded signal $x_{i}$ requires a search among $2^{B_{i}}$ quantization levels over $i=1,\dots,M$, complexity of the scheme is $\sum_{i=1}^{M}2^{B_{i}}$ instead of $2^{\sum_{i=1}^{M}B_{i}}$.
Hence, the order of complexity is the same as for PtPQ.

The scheme works as follows: for each RU $i$, the distance metric (\ref{eq:g_MQ}) is minimized by considering only the terms corresponding to the first $i$ RUs by fixing the quantized values $\hat{x}_{1},\dots,\hat{x}_{i-1}$ obtained at the previous iterations.
This yields the mapping
\be
\hat{x}_{i}=\arg\min_{\hat{x}_{i}\in\hat{\mathcal{X}}_{i}}\sum_{k=1}^{K}\alpha_{k}\left|\mathbf{h}_{k}^{i}{}^{H}\left(\mathbf{w}_{\mathbf{H},k}^{i}s_{k}-\hat{\mathbf{x}}^{i}\right)\right|^2,
\ee
where $\mathbf{a}^{i}$ represents the first $i$ elements of a vector $\mathbf{a}$, and in $\hat{\mathbf{x}}^{i}$ the first $(i-1)$ elements are fixed from the previous iterations.

\bfig [t]
\bc
\centering \epsfig{file=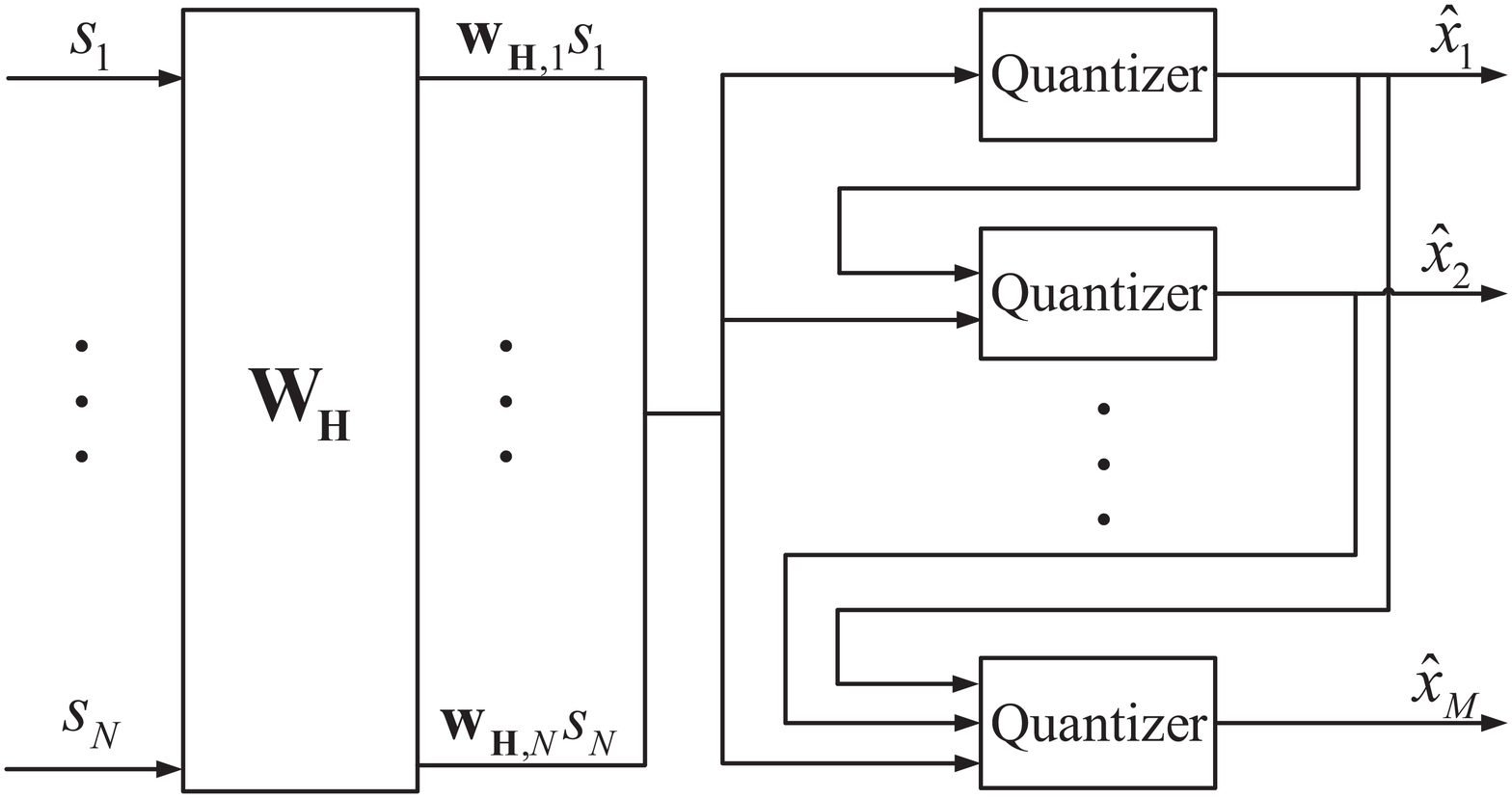, width=10cm,clip=}
\ec
\caption{Block diagram of the successive MQ implementation via sequential block quantizer steps ($d=1$).}
\label{fig:SMQ}
\efig

This approach can be generalized in order to offer a flexible solution that generates both the scheme defined above and MQ as discussed in Sec. \ref{ch:sub_MQ_fPC}.
To this end, successive MQ implementation can operate over blocks of size $d\in[1,M]$ at each step. 
In particular, the precoded vector $[x_{1+(i-1)d},\dots,x_{id}]$ is mapped to $[\hat{x}_{1+(i-1)d},\dots,\hat{x}_{id}]$ as 
\be
[\hat{x}_{1+(i-1)d},\dots,\hat{x}_{id}]=\arg\min_{[\hat{x}_{1+(i-1)d},\dots,\hat{x}_{id}]\in\hat{\mathcal{X}}_{1+(i-1)d}\times\cdots\times\hat{\mathcal{X}}_{id}}\sum_{k=1}^{K}\alpha_{k}\left|\mathbf{h}_{k}^{id}{}^{H}\left(\mathbf{w}_{\mathbf{H},k}^{id}s_{k}-\hat{\mathbf{x}}^{id}\right)\right|^2
\ee
for $i=1,\dots,M/d$, where we assume $M/d$ to be an integer for simplicity.
The computational complexity of the scheme is $\sum_{i=1}^{M/d}2^{\sum_{i^{'}=1+(i-1)d}^{id}B_{i^{'}}}$.
Note that successive block MQ scheme reduces to conventional MQ for $d=M$ and to successive MQ via sequential PtPQ for $d=1$.

\section{Entropy-Constrained Multivariate Quantization} \label{ch:MQ_VL}
In the previous sections, we have studied symbol-by-symbol quantization, which produces a fixed-length description, i.e., the same number of bits, for every baseband sample.
These descriptions may be redundant, and hence in this section, we consider the practically relevant case in which the quantizer is followed by a separate entropy encoder, or compressor, for each fronthaul link that produces variable-length descriptions for each sample.
The variable-rate compressor assigns more bits to the most used quantization levels and less bits to the least used levels, so that the average quantization output is $B$ bit/symbol.
The resulting quantization-compression system has the advantage of potentially reducing the fronthaul overhead for a given quantization resolution,
although this gain comes at the price of requiring the implementation of a buffer per fronthaul link in order to smooth out the variance of the variable-length descriptions \cite{Gray}.

We pursue here the optimization of the discussed quantization-compression system with both PtPQ and MQ by adopting the standard framework of entropy-constrained optimization \cite{Gray}.
We focus on the separate optimization of precoding and quantization in order to simplify the treatment, but the discussion could be extended to the joint optimization of precoding and quantization by following similar steps as in the previous section.

\subsection{Conventional Point-to-Point Quantization}
Following the entropy-constrained quantization design, for PtPQ, we modify the criterion (\ref{eq:opt_SQ}) by adding a penalty proportional to the entropy of the quantized signal \cite{Chou}.
We allow for a codebook of size $2^{B_{i}^{'}}$ with $B_{i}^{'}\geq B_{i}$ in order to leverage the larger resolution afforded by entropy coding.
This yields the problem of minimizing the Lagrangian
\be \label{eq:obj_ECSQ}
\min_{\hat{\mathcal{X}}_{i}=\{\hat{x}_{i}^{(j)}\}_{j=1}^{2^{B_{i}^{'}}}}E\left[\left(x_{i}-\hat{x}_{i}^{(f_{\hat{\mathcal{X}}_{i}}(x_{i}))}\right)^2\right]+\lambda H\left(\hat{x}_{i}^{(f_{\hat{\mathcal{X}}_{i}}(x_{i}))}\right),
\ee
where $\lambda$ is a weight (or Lagrangian multiplier), whose value determines a particular rate-distortion tradeoff,
$f_{\hat{\mathcal{X}}}(\cdot)$ is the quantization mapping to be optimized based on (\ref{eq:obj_ECSQ}), and $H(\cdot)$ is the entropy of the argument.

Following \cite{Chou}, an iterative algorithm that tackles the optimization of the codebooks as per criterion (\ref{eq:obj_ECSQ}) is summarized in Algorithm \ref{al:SQ_EC}.
In this algorithm, function
\be \label{eq:g_ecSQ}
g_{\hat{\mathcal{X}}_{i},\mathcal{S}_{i}}(x_{i})= \arg\min_{j}\left(x_{i}-\hat{x}_{i}^{(j)}\right)^2-\lambda \log_{2}p(\hat{x}_{i}^{(j)}),
\ee
is defined, where ties are arbitrarily broken; $\mathcal{S}_{i}=\{\mathcal{S}_{i}^{(j)}\}_{j=1}^{2^{B_{i}^{'}}}$ is the set of minimum-distortion partitions of the training samples;
and $p(\hat{x}_{i}^{(j)})=|\mathcal{S}_{i}^{(j)}|/N_{s}$ is the fraction of the training samples that are mapped to the index $j$ for RU $i$.
Moreover, in Algorithm \ref{al:SQ_EC}, the entropy of the quantized output is computed as
\be \label{eq:ent_SQ}
H\left(\hat{x}_{i}^{(f_{\hat{\mathcal{X}}_{i}}(x_{i}))}\right)=-\sum_{j=1}^{2^{B_{i}^{'}}}p(\hat{x}_{i}^{(j)})\log_{2}p(\hat{x}_{i}^{(j)}).
\ee

\begin{algorithm}[t]
  \caption{Entropy-constrained Point-to-Point Quantization (PtPQ) Design}
  \begin{algorithmic}
    \State \textbf{Input}: Generate $N_{s}$ training values $\mathcal{S}=\{\mathbf{s}(1),\dots,\mathbf{s}(N_{s})\}$. Set outer loop threshold $\tau \geq 0$ and inner loop threshold $\epsilon \geq 0$.
    \State \textbf{Initialization}: Set lower and upper bounds of Lagrange multiplier as $\lambda^{l}$ and $\lambda^{u}$, respectively. Initialize the Lagrange multiplier $\lambda=\lambda^{l}$.
    \Repeat
    \State $\bullet$ Initialize the codebook as $\hat{\mathcal{X}}_{i}^{[1]}$ with the number $2^{B_{i}^{'}}$ of quantization levels. Set $t=1$ and $D^{[0]}=\infty$.
    \Repeat
    \State \parbox[t]{\dimexpr\linewidth-\algorithmicindent-\algorithmicindent}{$\bullet$ Find the minimum-distortion partition of $\mathcal{S}$ as $\mathcal{S}_{i}^{(j)}=\{\mathbf{s}\in \mathcal{S}:g_{\hat{\mathcal{X}}_{i}^{[t]},\mathcal{S}_{i}^{[t]}}(\mathbf{w}_{\mathbf{H},i}\mathbf{s})=j\}$, where function $g_{\hat{\mathcal{X}}_{i}^{[t]},\mathcal{S}_{i}^{[t]}}$ is defined in (\ref{eq:g_ecSQ}) with $j=1,\dots,2^{B_{i}^{'}}$.}
    \State \parbox[t]{\dimexpr\linewidth-\algorithmicindent-\algorithmicindent}{$\bullet$ Compute the average distortion $D^{[t]}=1/N_{s}\sum_{n=1}^{N_{s}}(\mathbf{w}_{\mathbf{H},i}\mathbf{s}(n)-\hat{x}_{i}^{(g_{\hat{\mathcal{X}}_{i}^{[t]},\mathcal{S}_{i}^{[t]}}(\mathbf{w}_{\mathbf{H},i}\mathbf{s}(n)))})^2$.}
    \State \parbox[t]{\dimexpr\linewidth-\algorithmicindent}{$\bullet$ Obtain the new codebook as $\hat{\mathcal{X}}_{i}^{[t+1]}=\{\hat{x}_{i}^{(j)}:j=1,\dots,2^{B_{i}^{'}}\}$ by solving the quadratic convex problem (\ref{eq:Up_cb_SQ}).}
    \State $\bullet$ Set $t=t+1$.
    \Until {$(D^{[t-1]}-D^{[t]})/D^{[t]}\leq \epsilon$}.
    \State $\bullet$ Set $f_{\hat{\mathcal{X}}_{i}}(\cdot)$ to $g_{\hat{\mathcal{X}}_{i}^{[t]},\mathcal{S}_{i}^{[t]}}(\cdot)$.
    \State \parbox[t]{\dimexpr\linewidth-\algorithmicindent}{$\bullet$ Update $\lambda$ from bisectional search method: if $H(\hat{x}_{i}^{(f_{\hat{\mathcal{X}}_{i}}(x_{i}))})>B_{i}$, then $\lambda^{l}=(\lambda^{l}+\lambda^{u})/2$ and $\lambda=\lambda^{l}$; if $H(\hat{x}_{i}^{(f_{\hat{\mathcal{X}}_{i}}(x_{i}))})<B_{i}-\tau$, then $\lambda^{u}=(\lambda^{l}+\lambda^{u})/2$ and $\lambda=\lambda^{u}$.}
    \Until $H(\hat{x}_{i}^{(f_{\hat{\mathcal{X}}_{i}}(x_{i}))})\in[B_{i}-\tau,B_{i}]$.
    \State \textbf{Output}: Codebook $\hat{\mathcal{X}}_{i}=\hat{\mathcal{X}}_{i}^{[t]}$.
  \end{algorithmic}
  \label{al:SQ_EC}
\end{algorithm}

\subsection{Multivariate Quantization}
Similar to PtPQ, for MQ, we modify the codebook design problem (\ref{eq:wmse}) as the minimization of
\be \label{eq:obj_ECMQ}
\sum_{k=1}^{N}\alpha_{k}E\left[\left|\mathbf{h}_{k}^{H}\left(\mathbf{x}-\hat{\mathbf{x}}^{(f_{\hat{\mathcal{X}}}(\mathbf{x}))}\right)\right|^2\right]+ \sum_{i=1}^{M}\lambda_{i}H\left(\hat{x}_{i}^{(f_{\hat{\mathcal{X}}_{i}}({x}_{i}))}\right),
\ee
where $f_{\hat{\mathcal{X}}}(\cdot)$ is a mapper to be optimized based on (\ref{eq:obj_ECMQ}) over the quantization codebook $\hat{\mathcal{X}}=\hat{\mathcal{X}}_{1}\times\cdots\times \hat{\mathcal{X}}_{M}$, which consists of the Cartesian product of the sets $\hat{\mathcal{X}}_{i}=\{\hat{x}_{i}^{(1)},\dots,\hat{x}_{i}^{(2^{B_{i}^{'}})}\}$ of the quantization levels $\hat{x}_{i}^{(j)}$ for each RU $i$.
The resulting algorithm, which follows the approach in \cite{Chou}, is summarized in Algorithm \ref{al:MQ_EC}.
In the algorithm, we have defined the function
\be \label{eq:g_ecMQ}
g_{\hat{\mathcal{X}},\mathbf{H},\mathcal{S}}(\mathbf{x}) = \arg\min_{j_{1},\dots,j_{M}} \sum_{k=1}^{N}\alpha_{k} \left|\mathbf{h}_{k}^{H}\left(\mathbf{w}_{\mathbf{H},k}s_{k}-\hat{\mathbf{x}}^{(j_{1},\dots,j_{M})}\right)\right|^{2}-\sum_{i=1}^{M}\lambda_{i}\log_{2}p(\hat{x}_{i}^{(j_{i})}),
\ee
\sloppy{where $\mathcal{S}=\{\mathcal{S}^{(j_{1},\dots,j_{M})}\}_{i=1}^{M}{}_{j_{i}=1}^{2^{B_{i}^{'}}}$ is the set of minimum-distortion partitions of training samples and $\hat{\mathbf{x}}^{(j_{1},\dots,j_{M})}=[\hat{x}_{1}^{(j_{1})},\dots,\hat{x}_{M}^{(j_{M})}]$ is taken from the codebook $\hat{\mathcal{X}}$.}

\begin{algorithm} [t]
  \caption{Entropy-constrained Multivariate Quantization (MQ) Design}
  \begin{algorithmic}
    \State \textbf{Input}: Generate $N_{h}$ independent training channels $\mathcal{H}=\{\mathbf{H}(1),\dots,\mathbf{H}({N_{h}})\}$ from the known channel distribution and $N_{s}$ independent training values $\mathcal{S}=\{\mathbf{s}(1),\dots,\mathbf{s}(N_{s})\}$ from the known distribution of $\mathbf{s}$. Set outer loop threshold $\tau \geq 0$ and inner loop threshold $\epsilon \geq 0$.
    \State \textbf{Initialization}: For all $i=1,\dots,M$, set lower and upper bounds of Lagrange multiplier as $\lambda_{i}^{l}$ and $\lambda_{i}^{u}$, respectively. Initialize the $M \times 1$ Lagrange multiplier vector $\mathbf{\lambda}=[\lambda_{1},\dots,\lambda_{M}]^{T}=[\lambda_{1}^{l},\dots,\lambda_{M}^{l}]^{T}$.
    \Repeat
    \State $\bullet$ Initialize the codebook as $\hat{\mathcal{X}}^{[1]}$. Set $t=1$ and $D^{[0]}=\infty$.
    \Repeat
    \For {$m=1$ to $N_{h}$}
        \State \parbox[t]{\dimexpr\linewidth-\algorithmicindent-\algorithmicindent-\algorithmicindent\relax}{$\bullet$ Find the minimum-distortion partition of $\mathcal{S}$ when the channel is $\mathbf{H}(m)$ as $\mathcal{S}^{(j_{1},\dots,j_{M},m)}=\{\mathbf{s}\in \mathcal{S}: g_{\hat{\mathcal{X}}^{[t]},\mathbf{H}(m),\mathcal{S}^{[t]}}(\mathbf{W}_{\mathbf{H}(m)}\mathbf{s})=[j_{1},\dots,j_{M}]^T\}$ with $j_{i}=1,\dots,2^{B_{i}^{'}}$, $i=1,\dots,M$ and $m=1,\dots,N_{h}$, where function $g_{\hat{\mathcal{X}}^{[t]},\mathbf{H}(m),\mathcal{S}^{[t]}}$ is defined in (\ref{eq:g_ecMQ}).}
    \EndFor
    \State \parbox[t]{\dimexpr\linewidth-\algorithmicindent-\algorithmicindent\relax}{$\bullet$ \hspace{-6mm} Compute the average distortion $D^{[t]}=1/(N_{h}N_{s})\sum_{m=1}^{N_{h}}\sum_{n=1}^{N_{s}}\sum_{k=1}^{N}|\mathbf{h}_{k}(m)^{H}(\mathbf{w}_{\mathbf{H}(m),k}s_{k}(n)-\hat{\mathbf{x}}^{(g_{\hat{\mathcal{X}}^{[t]},\mathbf{H}(m),\mathcal{S}^{[t]}}(\mathbf{W}_{\mathbf{H}(m)}\mathbf{s}(n)))})|^2$.}
    \State \parbox[t]{\dimexpr\linewidth-\algorithmicindent}{$\bullet$ Obtain the updated codebook as $\hat{\mathcal{X}}^{[t+1]}$ by solving the quadratic convex problem (\ref{eq:Up_codebook}).}
    \State $\bullet$ Set $t=t+1$.
    \Until {$(D^{[t-1]}-D^{[t]})/D^{[t]}\leq \epsilon$}.
    \State $\bullet$ Set $f_{\hat{\mathcal{X}}}(\cdot)$ to $g_{\hat{\mathcal{X}}^{[t]},\mathbf{H}(m),\mathcal{S}^{[t]}}(\cdot)$.
    \State \parbox[t]{\dimexpr\linewidth-\algorithmicindent}{$\bullet$ Update $\mathbf{\lambda}$ from bisectional search method: for $i=1,\dots,M$, if $H(\hat{x}_{i}^{(f_{\hat{\mathcal{X}}_{i}}(x_{i}))})>B_{i}$, then $\lambda_{i}^{l}=(\lambda_{i}^{l}+\lambda_{i}^{u})/2$ and $\lambda_{i}=\lambda_{i}^{l}$; if $H(\hat{x}_{i}^{(f_{\hat{\mathcal{X}}_{i}}(x_{i}))})<B_{i}-\tau$, then $\lambda_{i}^{u}=(\lambda_{i}^{l}+\lambda_{i}^{u})/2$ and $\lambda_{i}=\lambda_{i}^{u}$.}
    \Until $H(\hat{x}_{i}^{(f_{\hat{\mathcal{X}}_{i}}(x_{i}))})\in[B_{i}-\tau,B_{i}]$ for all $i=1,\dots,M$.
    \State \textbf{Output}: Codebook $\hat{\mathcal{X}}=\hat{\mathcal{X}}^{[t]}$.
  \end{algorithmic}
  \label{al:MQ_EC}
\end{algorithm}

\section{Numerical Results} \label{ch:NR}
Throughout this section, we assume that every RU is subject to the same power constraint $P$ and has equal fronthaul capacity $B_{1}=\cdots=B_{M}=B$ bit/symbol.
The channel model follows \cite{Adhikary}, so that the channel vector is distributed as $\mathbf{h}_{k}\sim \mathcal{CN}(\mathbf{0},\mathbf{R}_{k})$, where the channel correlation matrix follows the one-ring scattering model.
With $\lambda/2$-spaced uniform linear arrays of RUs, we specifically have $\mathbf{R}_{k}=\mathbf{R}_{k}(\theta_{k},\Delta_{k})$ for an UE $k$ located
at a relative angle of arrival $\theta_{k}$ with angular spread $\Delta_{k}$, where the element $(m,n)$ of matrix $\mathbf{R}_{k}(\theta_{k},\Delta_{k})$ is given by
\be \label{eq:channel_model}
\left[\mathbf{R}_{k}(\theta_{k},\Delta_{k})\right]_{m,n}=\frac{1}{2\Delta_{k}}\int_{\theta_{k}-\Delta_{k}}^{\theta_{k}+\Delta_{k}}\exp^{-j\pi(m-n)\sin(\phi)}d\phi.
\ee
We set $\theta_{k}$ equal to $\pi/4$ and the angular spread $\Delta_{k}$ is equal to $2\pi$, except for Fig. \ref{fig:RvsBdel}.
Note that, with this choice of $\Delta_{k}$, matrix $\mathbf{R}_{k}$ is full rank.
For all design schemes, we assume Gaussian training samples for $\mathbf{s}$, and we set $\epsilon=0.001$ in Algorithms \ref{al:SQ}, \ref{al:MQ}, \ref{al:MQ_MM}, \ref{al:SQ_EC}, and \ref{al:MQ_EC}, and $r_{\max}=5$ in Algorithm \ref{al:DC}.

\bfig [t]
\bc
\centering \epsfig{file=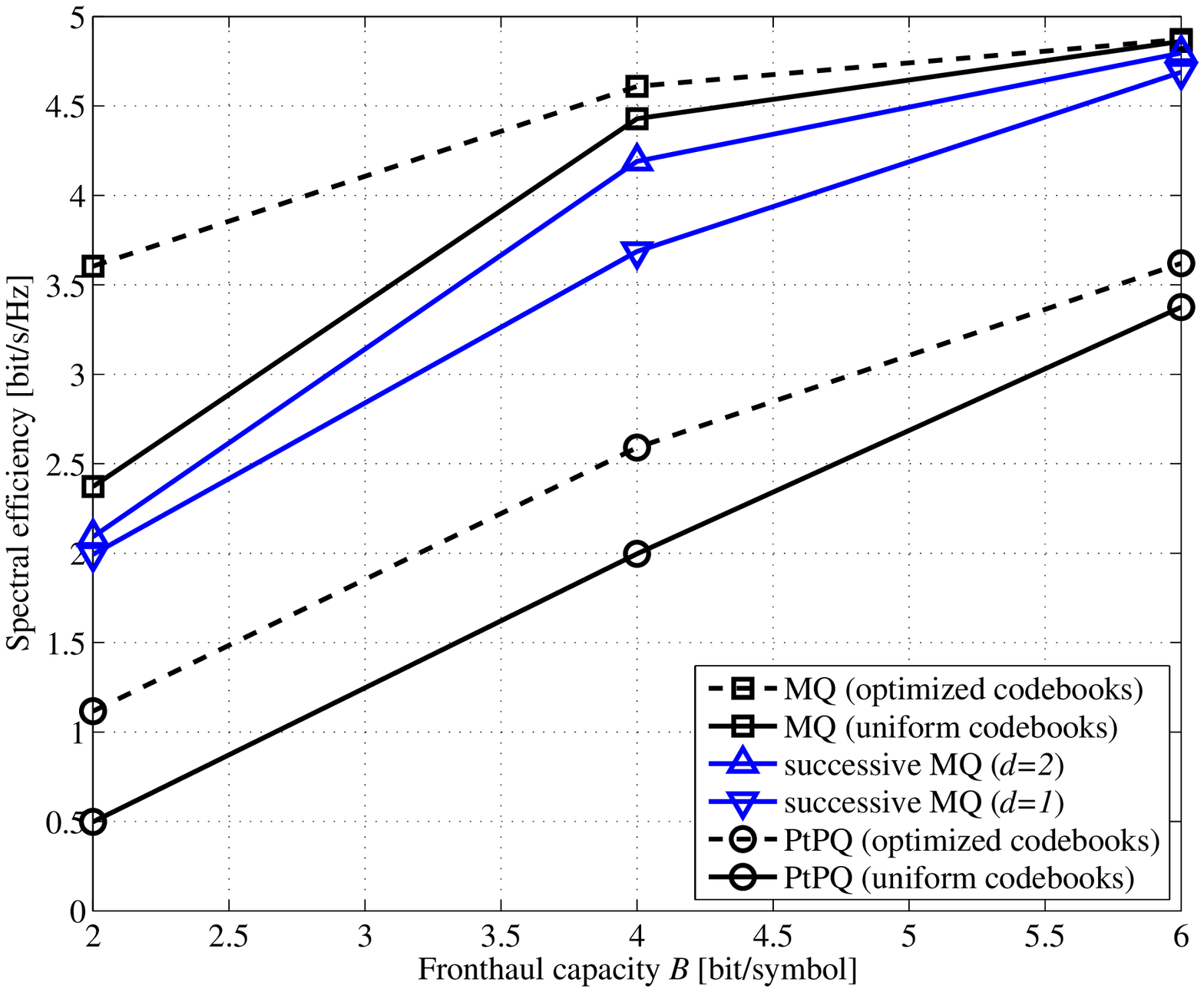, width=11cm,clip=}
\ec
\vspace{-5mm}
\caption{Spectral efficiency $R$ versus the fronthaul capacity $B$ of MQ and PtPQ with/without optimized codebook ($M=4$, $N=1$, $P=10$ dB, and $\gamma$=0.2).}
\label{fig:RvsB_cb}
\efig

We first compare the performance of PtPQ and MQ in the presence of: (\textit{i}) fixed codebooks, whose quantization levels are chosen to be uniformly spaced and whose dynamic range is based on training samples for both channel and symbols, as elaborated on in Sec. \ref{ch:MQ_fPC}; (\textit{ii}) optimized codebooks as discussed in Sec. \ref{ch:MQ_sep}; (\textit{iii}) reduced-complexity successive block MQ with uniform codebooks as studied in Sec. \ref{ch:sbMQ}.
We assume the separate optimization of quantization and precoding, which is fixed to the matched beamformer.
We consider a C-RAN with a single user ($N=1$), $M=4$ RUs, $P=10$ dB.

Fig. \ref{fig:RvsB_cb} shows the spectral efficiency, computed as $R=\sum_{k=1}^{N}\log_{2}(1+\textrm{SNR}_{k}^{\mathrm{eff}})$ bits/s/Hz, with the effective SNR given in (\ref{eq:effSNR}), versus the fronthaul capacity $B$.
We observe the significant gains of MQ over PtPQ, and also the relevant performance benefits of codebook optimization, particularly in the regime of low fronthaul capacity.
In contrast, it is seen that the performance of both PtPQ and MQ with fixed codebooks converge to that with optimized codebooks in the regime of a large enough fronthaul capacity.
Finally, with $d=1$, successive MQ outperforms PtPQ while having the same complexity order, and the performance can be significantly improved with $d=2$.
We recall that successive MQ with $d=4$ coincides with MQ. 

We then aim at assessing the advantages of MQ over PtPQ and of joint over separate optimization.
For separate optimization, following the approach in \cite{Park:TSP}, we first optimize precoding only by using Algorithm \ref{al:DC} with $\mathbf{\Omega}_{\mathbf{H}}=\mathbf{0}$ and with the per-antenna power constraint modified to $\mathbf{e}_{i}^{T}(\sum_{k=1}^{N}\mathbf{V}_{\mathbf{H},k}+\mathbf{\Omega}_{\mathbf{H}})\mathbf{e}_{i} \leq \gamma$ for some parameter $\gamma \in [0,1]$;
and then optimize quantization only by using Algorithm \ref{al:SQ} for PtPQ and Algorithm \ref{al:MQ} for MQ.
As in \cite{Park:TSP}, the parameter $\gamma$ defines the margin $(1-\gamma)$ of transmit power that is used to accommodate the presence of quantization noise in the transmitted signal.
Instead, for joint optimization, we implement Algorithm \ref{al:MQ_MM} and the corresponding scheme for PtPQ as discussed in Sec. \ref{ch:MQ_JD}.
We also consider as reference the performance obtained with no fronthaul capacity limitation, which amounts to adopting Algorithm \ref{al:DC} with $\mathbf{\Omega}_{\mathbf{H}}=\mathbf{0}$.

\bfig [t]
\bc
\centering \epsfig{file=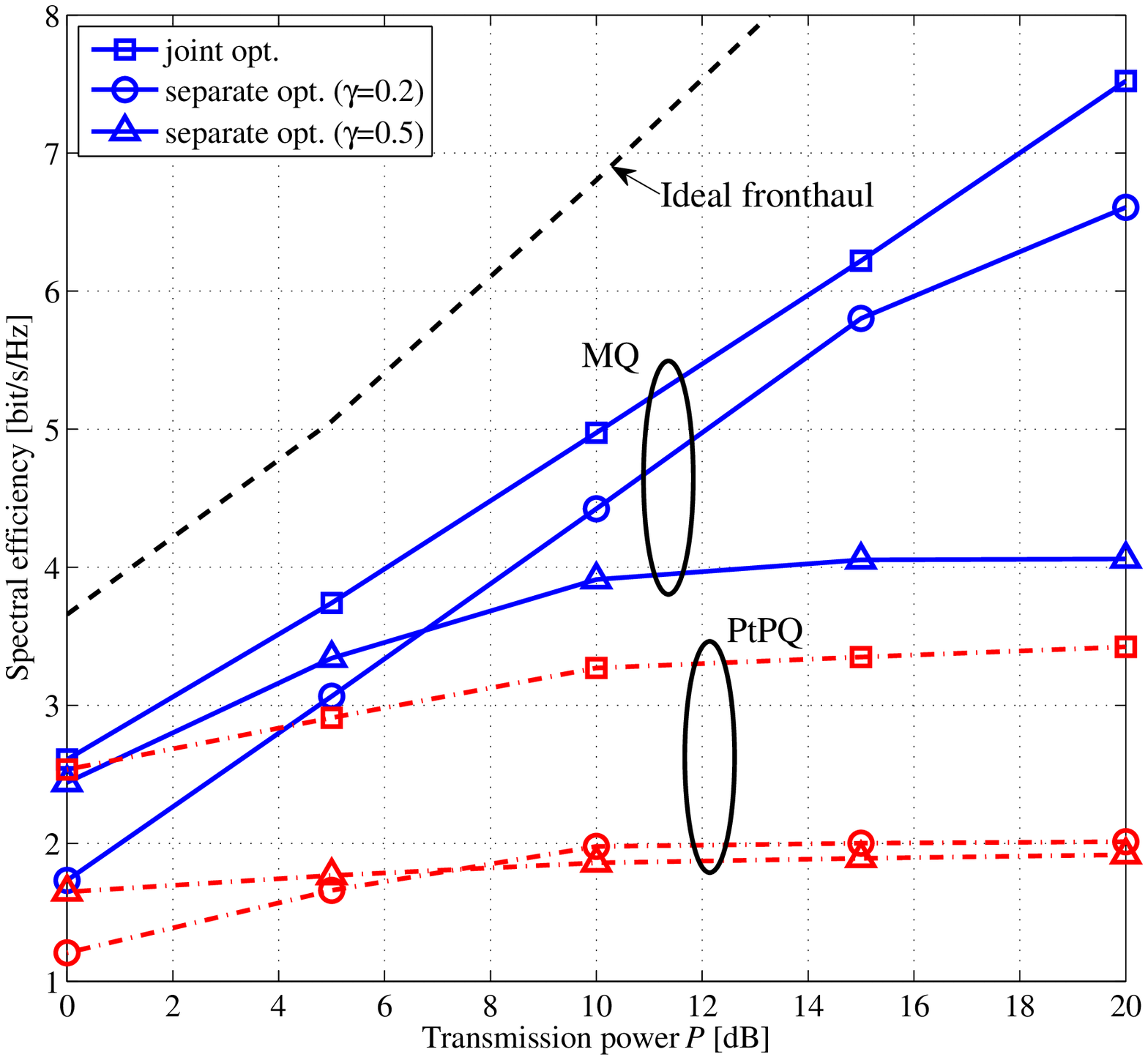, width=11cm,clip=}
\ec
\vspace{-5mm}
\caption{Spectral efficiency $R$ versus the transmission power $P$ for separate and joint design of MQ and PtPQ ($M=4$, $N=1$, and $B=3$ bit/symbol).}
\label{fig:RvsP}
\efig

We now show the spectral efficiency varying with the transmit power $P$ in Fig. \ref{fig:RvsP} with $M=4$ RUs, $N=1$ UE, and $B=3$ bit/symbol.
It is seen that the MQ significantly outperforms PtPQ, particulary for high SNR $P$.
This is because the gain of a more sophisticated quantization strategy is more pronounced in the high SNR regime in which the distortion caused by quantization becomes the dominant factor.
We also observe the significant advantages of joint optimization with respect to separate optimization and the sensitivity of the latter to the choice of the power offset parameter $\gamma$.
The performance gains of joint optimization can be ascribed to the capability to jointly select the signal and quantization noise spatial properties as discussed in Sec. \ref{ch:MQ_JD}.

\bfig [t]
\bc
\centering \epsfig{file=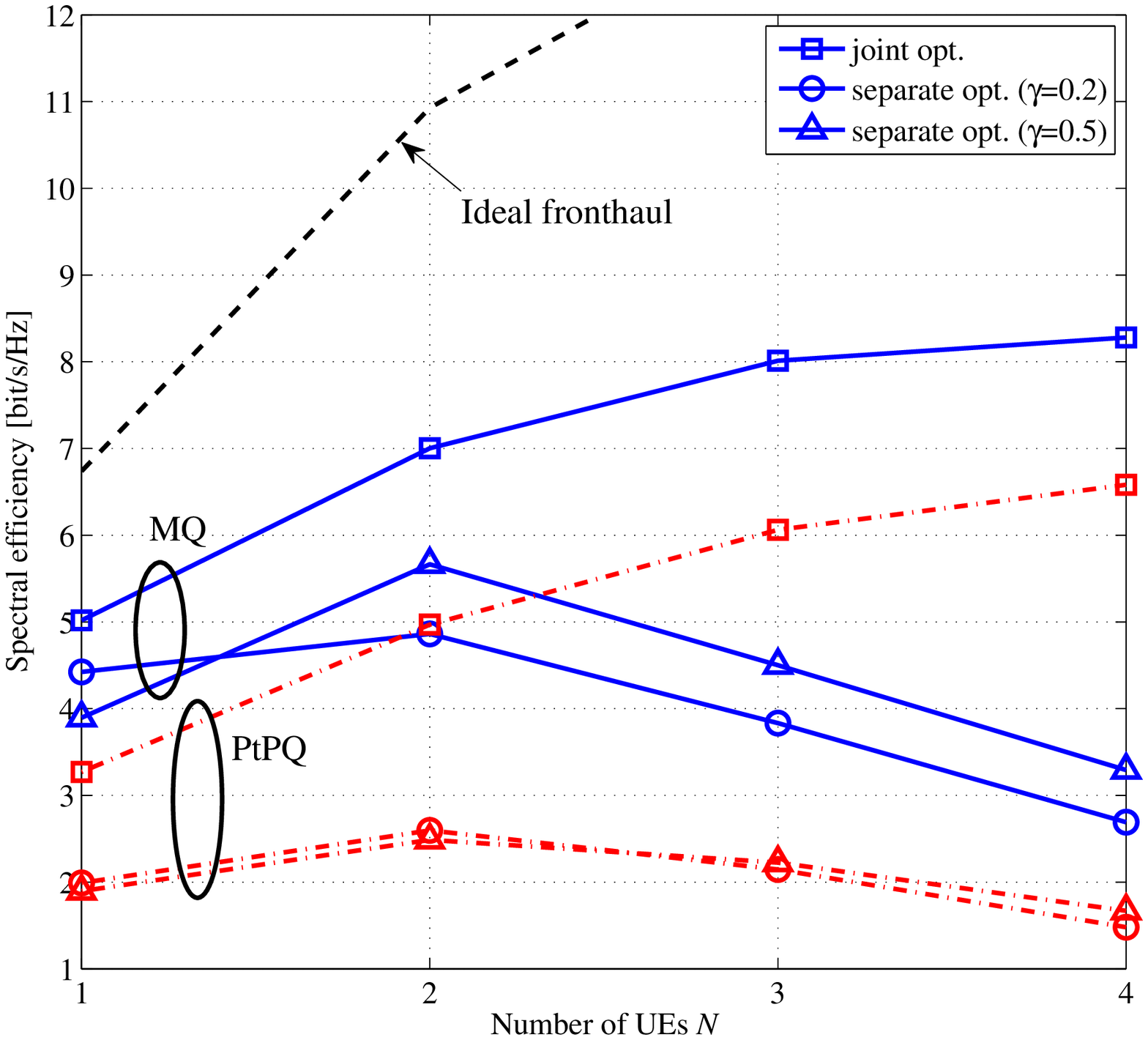, width=11cm,clip=}
\ec
\vspace{-5mm}
\caption{Spectral efficiency $R$ versus the number of UEs $N$ for separate and joint design of MQ and PtPQ ($M=4$, $B=3$ bit/symbol, and $P=10$ dB).}
\label{fig:RvsN}
\efig

In Fig. \ref{fig:RvsN}, we fix $M=4$, $P=10$ dB, and $B=3$ bit/symbol, and let the number $N$ of UEs increase.
We observe that the performance gains of MQ are realized across all values of $N$.
In particular, with joint design of codebook and precoding matrix, the performance gain of MQ is achieved around $2$ bit/s/Hz compared to PtPQ.

\bfig [t]
\bc
\centering \epsfig{file=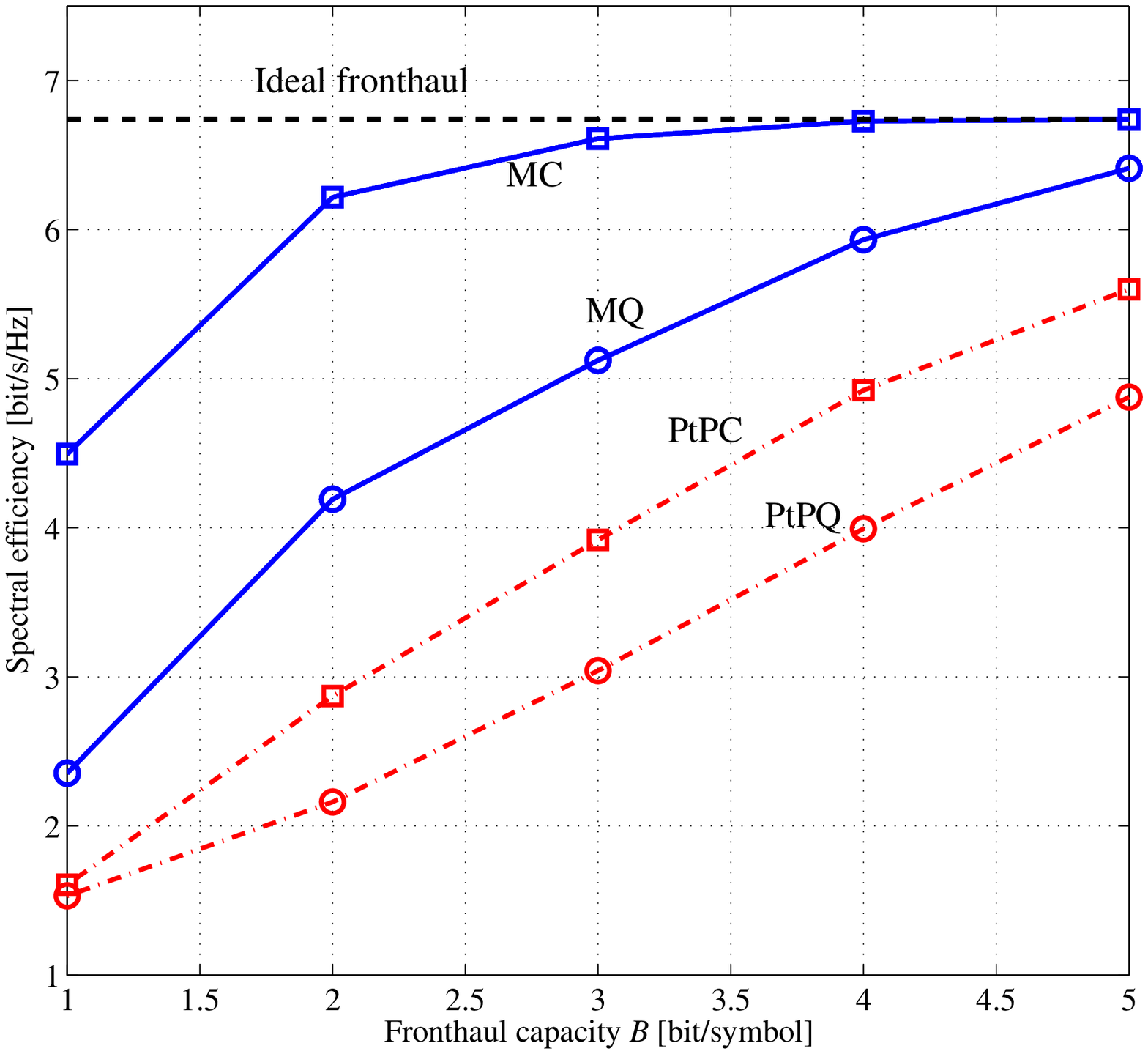, width=11cm,clip=}
\ec
\vspace{-5mm}
\caption{Spectral efficiency $R$ versus the fronthaul capacity $B$ for joint design of MQ and PtPQ compared to the corresponding performance of information-theoretically optimal block quantizers as studied in \cite{Park:TSP} ($M=4$, $N=1$, and $P=10$ dB).}
\label{fig:RvsB}
\efig

We then investigate the performance for varying fronthaul capacity $B$ in Fig. \ref{fig:RvsB}.
We again consider $M=4$, $N=1$, and $P=10$ dB.
Here, we compare symbol-by-symbol quantization as studied throughout this paper with information-theoretically optimal block quantizers as considered in \cite{Park:TSP}.
As per standard information-theoretic arguments, such quantizers operate over arbitrarily long blocks of symbols rather than on a per-symbol basis.
Their performance is evaluated here by using the joint optimization algorithm in \cite[Algorithm~1]{Park:TSP}.
For symbol-by-symbol PtPQ and MQ, we also assume joint optimization.
Fig. \ref{fig:RvsB} shows that, for sufficiently small fronthaul capacity $B$, block processing is able to significantly improve the performance of symbol-by-symbol quantization.
For instance, for $B=1$ bits/symbol, the performance of MQ in terms of spectral efficiency is approximately doubled. However, this performance gain decreases with $B$.
As an example, with $B=4$ bits/symbol, block processing provides only a gain of $0.8$ bit/s/Hz with respect to symbol by symbol quantization, demonstrating that a symbol-by-symbol approach can be an effective close-to-optimal solution as long as the fronthaul capacity is not too small.

\bfig [t]
\bc
\centering \epsfig{file=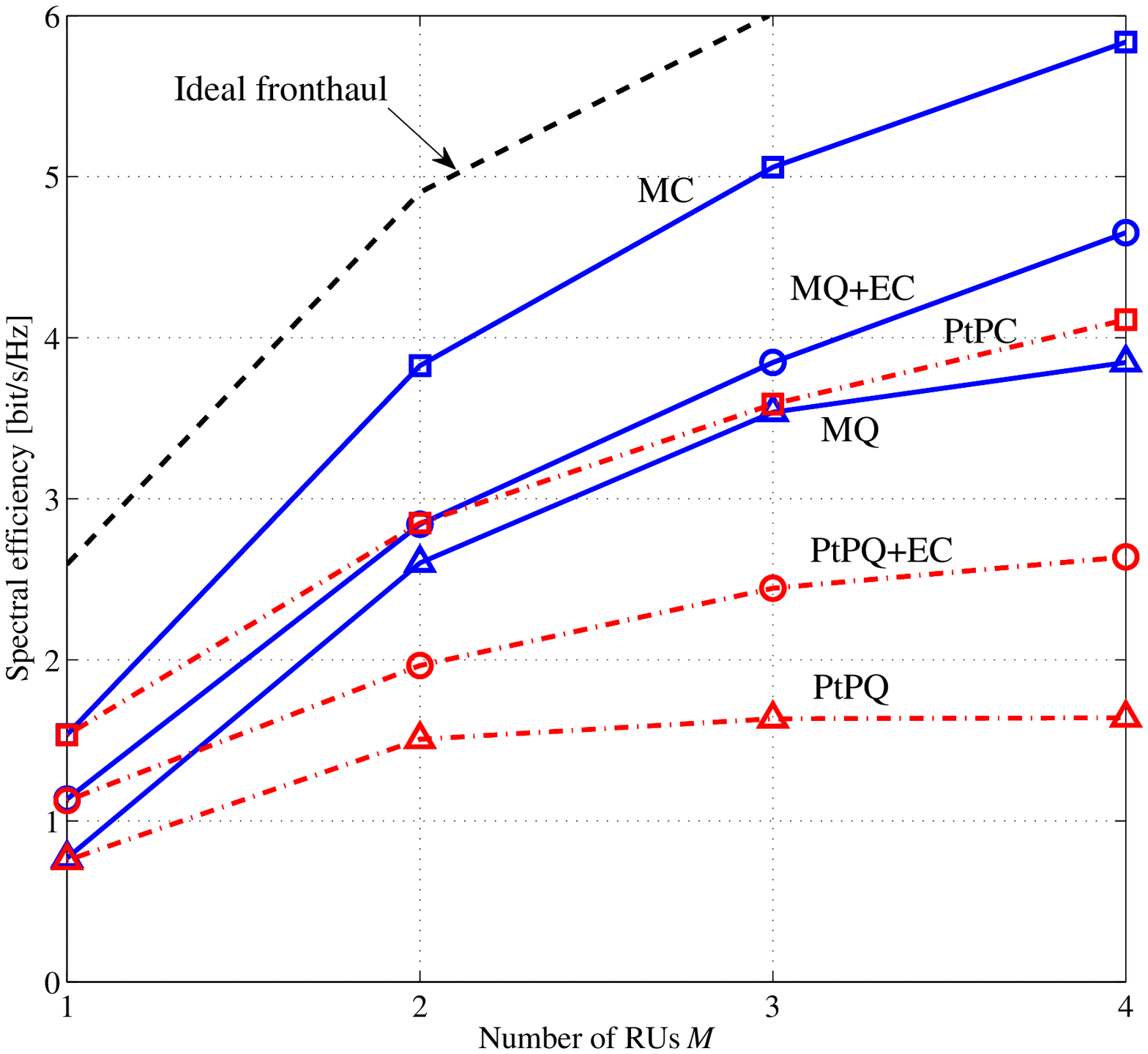, width=11cm,clip=}
\ec
\vspace{-5mm}
\caption{Spectral efficiency $R$ versus the number $M$ of RUs for PtPQ and MQ with entropy encoders following the quantizers and with optimal block quantizers ($N=1$, $P=10$ dB, $B=3$ bit/symbol, and $\gamma=0.5$).}
\label{fig:RvsM_blk_ec}
\efig

Fig. \ref{fig:RvsM_blk_ec} compares the spectral efficiency achievable by PtPQ and MQ, under both symbol-by-symbol and block processing, with the spectral efficiency achievable by the entropy-constrained PtPQ and MQ design when varying the number $M$ of RUs.
We set $N=1$, $P=10$ dB, $B=3$ bit/symbol, and $\gamma=0.5$.
In Algorithm \ref{al:SQ_EC} and \ref{al:MQ_EC}, we also select $\tau=0.05$, $B_{i}^{'}=B_{i}+1$, $\lambda^{l}=0$, and $\lambda^{u}=1.5$.
We first observe that the gain of MQ over PtPQ increases with a larger $M$ due to the increased number of degrees of freedom available for the design of the multivariate quantizer.
For instance, for $M=2$, MQ yields a spectral efficiency gains of $73\%$, while the gain increases to $143\%$ for $M=4$.
That is, as the number of RUs, and hence the dimension of the codebook, increases, MQ efficiently controls the spatial direction of the transmit signal.
As a general rule, block processing outperforms symbol-by-symbol quantization with entropy coding and the latter improves upon standard symbol-by-symbol quantization.
Moreover, it is observed that the relative gains of these three approaches are more significant for PtPQ than for MQ.
For instance, for $M=4$, in the case of PtPQ, entropy coding improves the spectral efficiency of standard quantization by $60\%$ and block processing improves the spectral efficiency by $160\%$.
Instead, for MQ, the corresponding gains are $21\%$ and $52\%$.
This can be interpreted as the effect of the capability of MQ to reduce the impact of the quantization error as compared to PtPQ,
hence making the use of more sophisticated compression techniques less relevant.
%Note that, in Fig. \ref{fig:RvsM_blk_ec}, this is particularly evident as the number of RUs increases given the enhanced effectiveness of MQ in this regime.

\bfig [t]
\bc
\centering \epsfig{file=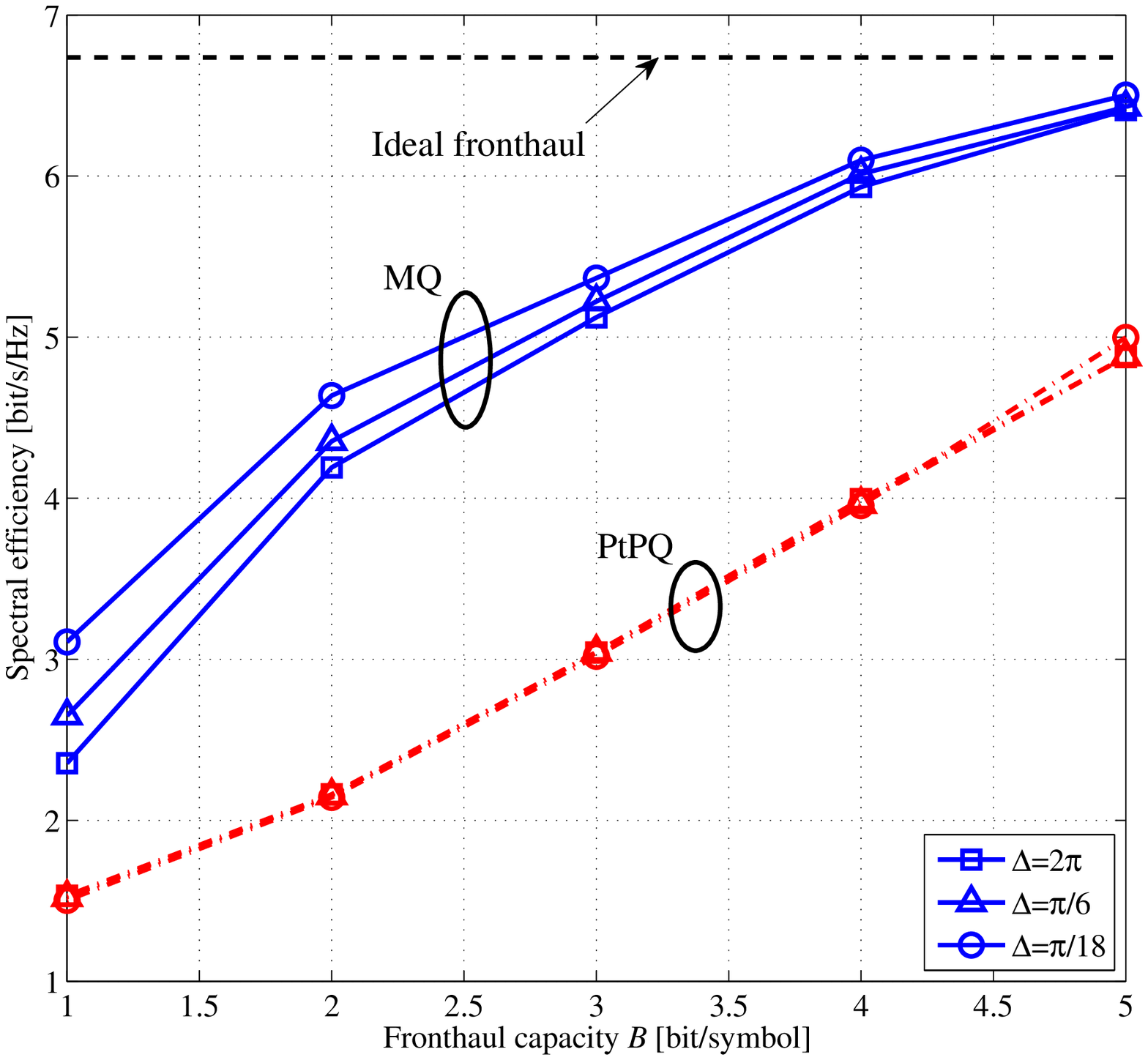, width=11cm,clip=}
\ec
\vspace{-5mm}
\caption{Spectral efficiency $R$ versus the fronthaul capacity $B$ for joint design of MQ and PtPQ varying with angular spread $\Delta$ ($M=4$, $N=1$, and $P=10$ dB).}
\label{fig:RvsBdel}
\efig

We finally turn to investigate the impact of angular spread $\Delta$ for the stochastic channel model (\ref{eq:channel_model}).
We consider $M=4$, $N=1$, and $P=10$ dB and plot the spectral efficiency as a function of fronthaul capacity $B$ for different values of $\Delta$ in Fig. \ref{fig:RvsBdel}.
It is seen that, while the performance of PtPQ does not depend on $\Delta$, MQ benefits from a smaller value of $\Delta$.
This is expected because MQ can take advantage of a lower-rank channel by properly designing the codebook based on long-term CSI.
Nevertheless, the spectral efficiency loss observed for a larger $\Delta$ is rather small, demonstrating the effectiveness of the joint mapping carried out by MQ based on current CSI, even in the presence of fixed codebooks.
Furthermore, this loss decreases for larger values of $B$ in accordance with the discussion around Fig. \ref{fig:RvsB_cb} regarding the reduced gain of codebook optimization as a function of the long-term CSI for increasing $B$.

\bfig [t]
\bc
\centering \epsfig{file=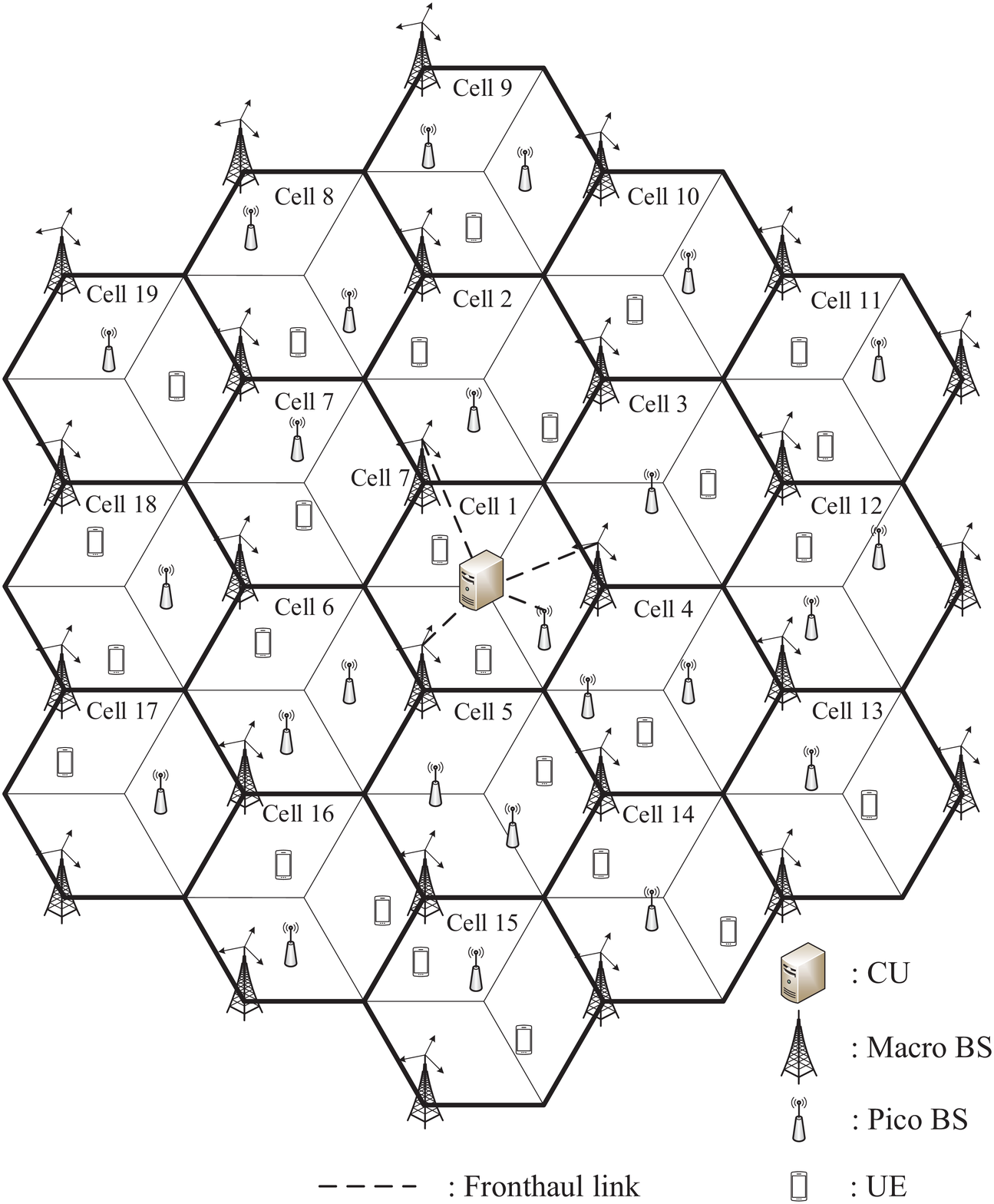, width=8.5cm,clip=}
\ec
\caption{Two-dimensional hexagonal cellular system with $19$ macro cells.
Each macro-BS has three sectorized antenna, while the pico-BSs and the UE have a single omni-directional antenna each.}
\label{fig:Cell}
\efig

\section{Performance under Standard Cellular Models} \label{ch:MQ_LTE}
In this section, we investigate the system-level performance of MQ over a cellular model specified in the LTE standard documents \cite{3gpp:2009}\cite{3gpp:2011}.
This model was also considered in \cite{Park:CISS} for the performance evaluation of MC.
A two-dimensional hexagonal cellular system with $19$ cells is assumed, where each macro-cell contains three fixed macro-base stations (BSs), a number of uniformly distributed pico-BSs, and $N$ uniformly distributed UEs (see Fig. \ref{fig:Cell}).
We assume the standard set-up in which each macro-BS has three sectorized antenna, while the pico-BSs and the UEs have a single omni-directional antenna each.
All the macro and pico-BSs in a given cell are connected to the same CU via orthogonal fronthaul links.
During $T$ time slots, we assume that the locations of pico-BSs and UEs are fixed and small-scale fading channel changes independently from slot to slot.
We also assume that the available bandwidth is partitioned into three bands with frequency reuse $F=1/3$ to minimize the inter-cluster interference, that is, cell $1$ suffers the interference from cells $8,10,12,14,16,18$.
The interference signals from other cells are treated as noise.
From \cite{3gpp:2009} and \cite{3gpp:2011}, the system parameters are summarized in Table \ref{tb:LTE} and we adopt the LTE spectral efficiency model to evaluate the system-level performance \cite[Annex~A]{3gpp:2009}, namely 
\be
R_{k}^{\textrm{LTE}} = \left\{
\begin{array}{ll}
0 &\textrm{if} ~ R_{k} \leq \log_{2}(1+10^{-1}), \\
\min\left(0.6R_{k},4.4\right) &\textrm{otherwise}.
\end{array}
\right.
\ee

As in \cite{Park:CISS}, we adopt the standard proportional-fair scheduler over the $T$ time slots.
The scheduler operates over each slot $t$ by adapting to the average rates $\bar{R}_{k,t}$ allocated to each UE $k$ in the previous time slots,
which is updated by $\bar{R}_{k,t}=\beta\bar{R}_{k,t-1}+(1-\beta)R_{k,t-1}$ with the forgetting factor $\beta\in[0,1]$.
To this end, we propose to first design the precoding matrix $\mathbf{W}_{\mathbf{H}}$ assuming zero quantization covariance matrix $\mathbf{\Omega}_{\mathbf{H}}=\mathbf{0}$ by solving the optimization problem (see, e.g., \cite{Beck})
\be \label{eq:LTE_opt}
\max_{\mathbf{W}_{\mathbf{H}}} &&\sum_{k=1}^{N}\frac{R_{k,t}}{\bar{R}_{k,t}^{\alpha}} \nonumber \\
\mathrm{s.t.}&&\|\mathbf{e}_{i}^{T}\mathbf{W}_{{\mathbf{H}}}\|^{2} \leq \gamma P_{i},
\ee
where $R_{k,t}$ is equal to the right-hand side in (\ref{eq:Rk}) with channel vectors corresponding to slot $t$ and $\alpha>0$ is a fairness constant.
The optimization problem (\ref{eq:LTE_opt}) can be solved using Algorithm \ref{al:DC}.
With the obtained precoding matrix $\mathbf{W}_{\mathbf{H}}$ from (\ref{eq:LTE_opt}), we then propose to apply MQ with a mapping function modified by the weights of the proportional-fair scheduler in a manner similar to (\ref{eq:LTE_opt}) as
\be \label{eq:LTE_mapping}
f_{\hat{\mathcal{X}},\mathbf{H}}(\mathbf{x})=\arg\min_{j_{1},\dots,j_{M}}\sum_{k=1}^{N}\left(\bar{R}_{k,t}^{\alpha}\right)^{-1}\left|\mathbf{h}_{k,t}^{H}\left(\mathbf{w}_{\mathbf{H},k}s_{k,t}-\hat{\mathbf{x}}^{(j_{1},\dots,j_{M})}\right)\right|^2.
\ee
The codebook $\hat{\mathcal{X}}$ can also be optimized by using the mapping function (\ref{eq:LTE_mapping}) with Algorithm \ref{al:MQ}.

\begin{table}[t] \bc
  \begin{tabular}{ | c | c | }
    \hline
    Parameters & Assumptions \\ \hline\hline
    System bandwidth & $10$ MHz \\ \hline
    Macro-BS path-loss & $\mathrm{PL}(\mathrm{dB})=128.1+37.6\log_{10}R$, $R:$ distance in kilometers \\ \hline
    Pico-BS path-loss & $\mathrm{PL}(\mathrm{dB})=38+30\log_{10}R$, $R:$ distance in kilometers \\ \hline
    Antenna pattern for sectorized macro-BS antennas & $A(\theta)=-\min(12(\theta/\theta_{3\mathrm{dB}})^{2},A_{m})$, $\theta_{3\mathrm{dB}}=65^{\circ}$, $A_{m}=20$ dB \\ \hline
    Log-normal shadowing (standard deviation) & $10$ dB (macro-BS), $6$ dB (pico-BS) \\ \hline
    Transmit power & $46$ dBm (macro-BS), $24$ dBm (pico-BS) \\ \hline
    Antenna gain after cable loss & $15$ dBi (macro-BS), $0$ dBi (pico-BS) \\ \hline
    Noise figure & $9$ dB (UE) \\ \hline
  \end{tabular}
  \ec \caption{The system parameters for the standard cellular model studied in Sec. \ref{ch:MQ_LTE}.} \label{tb:LTE}
\end{table}

In Fig. \ref{fig:LTE}, we plot the cell-edge throughput, which is defined as the $5\%$-ile spectral efficiency, versus the average spectral efficiency with one pico-BS, $N=4$ UEs, $(B_{\mathrm{macro}},B_{\mathrm{pico}})=(4,2)$ bit/symbol, where $B_{\mathrm{macro}}$ is the fronthaul capacity of each macro-BS and $B_{\mathrm{pico}}$ is the fronthaul capacity of pico-BS, $T=5$, and $\beta=0.5$.
The curves are obtained by varying the fairness constant $\alpha$ from $0.5$ to $2$.
In accordance with the fact that the fairness requirement becomes more pronounced as $\alpha$ gets larger, the $5\%$-ile spectral efficiency is seen to increased with $\alpha$ at the cost of a lower spectral efficiency.
Moreover, with an optimized codebook, we observe that MQ can achieve about $2\times$ gain in terms of edge-cell rate as compared to PtPQ in a manner similar to the results reported in \cite{Park:CISS} for MC over PtPC.
It is also seen that, although both of PtPQ and MQ with fixed uniform quantization codebook are not able to achieve the spectral efficiency larger than $0.8$ bps/Hz,
those with optimized codebook can achieve over than $1$ bps/Hz for $\alpha=0.5$.

\bfig [t]
\bc
\centering \epsfig{file=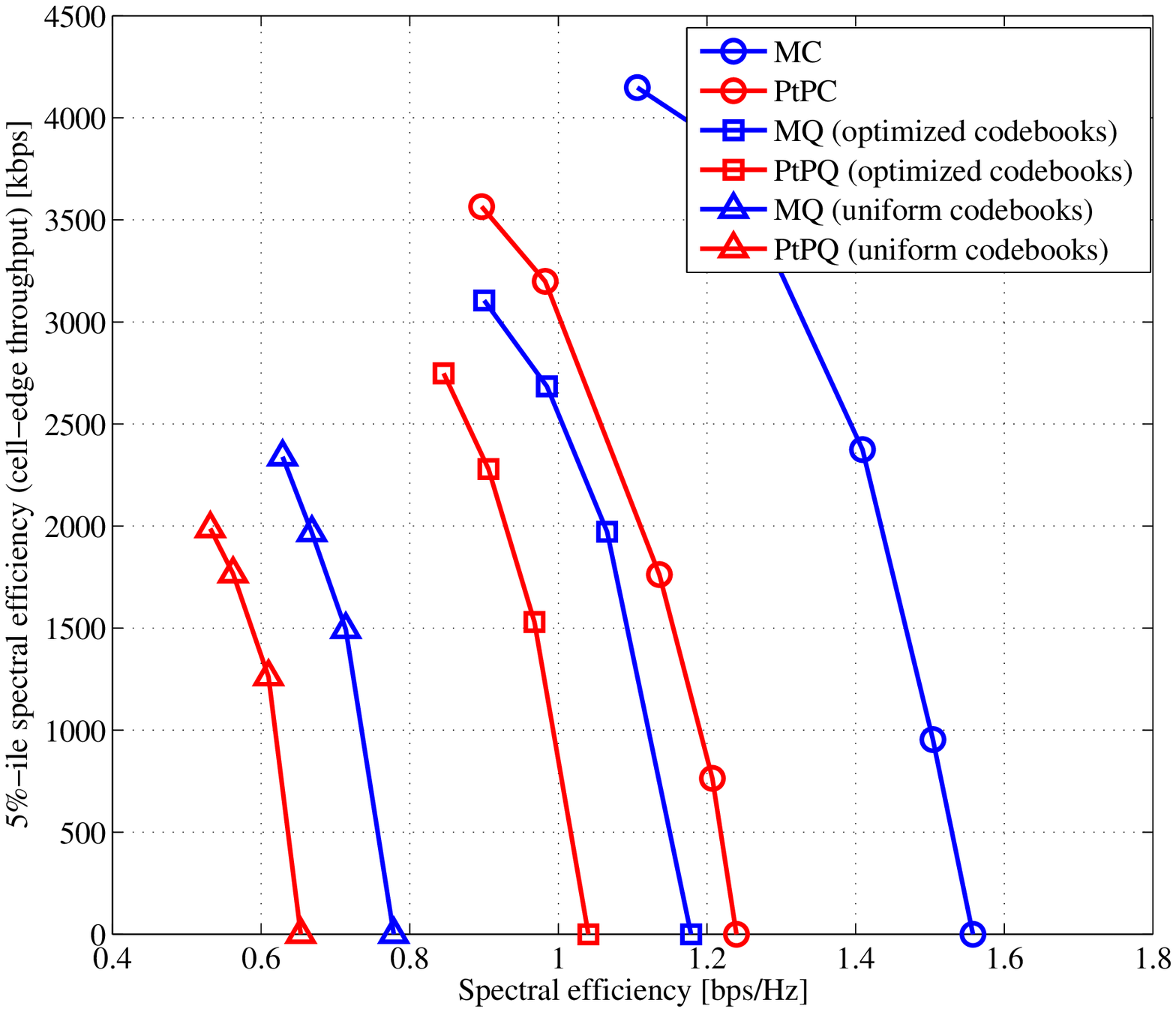, width=11cm,clip=}
\ec
\vspace{-5mm}
\caption{Cell-edge throughput, i.e., $5\%$-ile spectral efficiency, versus the average spectral efficiency for MC, PtPC, MQ, and PtPQ by varying the fairness constant $\alpha$ ($M=1$ pico-BS, $N=4$ UEs, $(B_{\mathrm{macro}},B_{\mathrm{pico}})=(4,2)$ bit/symbol, $T=5$, and $\beta=0.5$).}
\label{fig:LTE}
\efig

\section{Conclusions} \label{ch:Con}
This paper has presented a symbol-by-symbol implementation of the multivariate compression scheme proposed in \cite{Park:TSP} for the downlink of Cloud-Radio Access Network (C-RAN).
It has been demonstrated that the proposed Multivariate Quantization (MQ) scheme yields significant performance gain over per-fronthaul link Point-to-Point Quantization (PtPQ) as carried out in the Common Public Radio Interface (CPRI) standard, particularly when jointly designed with precoding.
This gain is realized without requiring any modification of the radio units.
Furthermore, we have proposed the reduced-complexity MQ by means of successive block quantization.
In turn, we have studied the additional potential benefits of following symbol-by-symbol quantization with variable-length compression.
As related future work, we observe that the approach introduced here could be extended to longer blocks by means of vector (structured) quantization \cite{Gray}, and that it could be combined with other point-to-point compression techniques,
such as filtering, per-block scaling, predictive quantization and lossless compression \cite{Samardzija}-\cite{Si}.

%X\section*{Acknowledgment}

\end{document}